\documentclass[apj,normalem]{emulateapj}

\usepackage{color}
\def\gsim{\ \raise 3pt \hbox{$>$} \kern -8.5pt \raise -2pt \hbox{$\sim$}\ }
\def\lsim{\ \raise 3pt \hbox{$<$} \kern -8.5pt \raise -2pt \hbox{$\sim$}\ }

\def\rhessi{{\textit{RHESSI}}}
\def\mw{{microwave}}
\def\Mw{{Microwave}}

\usepackage{ulem}
\begin{document}
\title{Narrowband Gyrosynchrotron Bursts: Probing Electron Acceleration in Solar Flares}
\author{Gregory D. Fleishman\altaffilmark{1}, Gelu M. Nita\altaffilmark{1},
Eduard P. Kontar\altaffilmark{2}, and Dale E. Gary\altaffilmark{1} }

\altaffiltext{1}{Center For Solar-Terrestrial Research, New Jersey Institute of Technology, Newark, NJ 07102}
\altaffiltext{2}{SUPA, School of Physics and Astronomy, University of Glasgow, G12 8QQ, United Kingdom}


\begin{abstract}
Recently, in a few case studies we demonstrated that gyrosynchrotron \mw\ emission can be detected directly from the acceleration region when the trapped electron component is insignificant. For the statistical study reported here, we
have identified events with steep (narrowband) microwave spectra that do not show a significant trapped component and at the same time show evidence of source uniformity, which simplifies the data analysis greatly. Initially, we identified a subset of more than 20 radio bursts with such narrow spectra, having low- and high-frequency spectral indices larger than 3 in absolute value. A steep low-frequency spectrum implies that the emission is nonthermal (for optically-thick thermal emission, the spectral index cannot be steeper than 2), and the source is reasonably dense and uniform. A steep high-frequency spectrum implies that no significant electron trapping occurs; otherwise a progressive spectral flattening would be observed. Roughly half of these radio bursts have RHESSI data, which allows for detailed, joint diagnostics of the source parameters and evolution. Based on an analysis of radio-to-X-ray spatial relationships, timing, and spectral fits, we conclude that the \mw\ emission in these narrowband bursts originates directly from the acceleration regions, which have relatively strong magnetic field, high density, and low temperature. In contrast, the thermal X-ray emission comes from a distinct loop with smaller magnetic field, lower density, but higher temperature. Therefore, these flares occurred likely due to interaction between two (or more) magnetic loops.
\end{abstract}

\section{Introduction}

Solar flares are observed throughout the electromagnetic spectrum with numerous space and ground-based instruments, which capture both thermal and nonthermal manifestations of the flares. Nevertheless, our knowledge on the energy storage and release, magnetic reconnection, partition between particle acceleration and plasma heating, and thermal plasma response, remains limited and fragmentary. For example, when a flare acceleration region is detected using primarily X-ray data, this acceleration region contains a somewhat dense thermal plasma, because the high density makes the detection of the source in X-ray range much easier. On the other hand, the detections made using microwave data \citep{Fl_etal_2011, Fl_etal_2013} reveal rather tenuous acceleration regions. In addition, the flares themselves demonstrate extreme diversity of the energy partitions and acceleration efficiencies---from entirely thermal \citep{Gary_Hurford_1989, Fl_etal_2015} to essentially acceleration-dominated  \citep{2010ApJ...714.1108K, Fl_etal_2011}. This calls for a systematic search for the detection of acceleration sites in more events, understanding those events, and determining if they fall into distinct sub-categories. Here we report such a systematic search and corresponding analysis with the emphasis on the \mw\ data augmented by other available context data, primarily---the \rhessi\ X-ray data.

Microwave emission in solar flares can originate from energetic electrons in either directly the acceleration region or after their propagation to other regions within the magnetic structure \citep[see a detailed discussion of these different components in][]{Fl_etal_2011}. Accelerated electrons can in addition be magnetically trapped and accumulated at the looptop \citep[e.g.,][]{Melnikov_1994, Meln_Magun_1998, Bastian_etal_1998, Lee_Gary_2000, 2000ApJ...531.1109L, Kundu_etal_2001, melnikov_etal_2002}. The accumulated electrons can fill in a coronal part of the flaring loop, where their energy spectrum often evolves toward an ever harder energy spectrum due to the energy-dependent Coulomb losses.
The gyrosynchrotron (GS) spectrum produced by such a source is expected to be broadband \citep[see][for a few vivid examples]{1994SoPh..152..409L}: from a few GHz, where the emission is often optically thick, to a hundred GHz and beyond. Such broadband emission favors its routine detection over a wide range of frequencies. Indeed, the broadband continuum
bursts are easy to detect by radio instruments observing at a fixed set of frequencies with large gaps between the spectral channels, such as the Radio Solar Telescope Network (RSTN) or Nobeyama Polarimeters \citep[NoRP;][]{1985PASJ...37..163N}. The radio spectrum normally has a broad peak around 10~GHz separating optically thin and thick regions of the radio spectrum. Occasionally, more than one peak can be observed \citep{1998ARA&A..36..131B, 1990ApJ...361..290G, 1989SoPh..120..351S, 1990SoPh..125..343S, 1975SoPh...44..155G, 2004ApJ...605..528N, Fl_etal_2013,2013SoPh..284..541G}. 
More recent observations at frequencies above 100~GHz indicated the presence of a new rising-with-frequency component \citep{2012snc..book...61K}. Although a large number of models is discussed in connection with this mysterious component, the exact mechanism of the emission remains unclear \citep[][]{2010ApJ...709L.127F,2014SoPh..289.3017Z,2014ApJ...791...31K}.

Although such broadband emission is rather common, it is not unique.  \citet{2006A&A...457..693G} described narrowband emission over a few seconds at the rise phase of a solar microwave burst. Our recent studies \citep{Fl_etal_2011,Fl_etal_2013} have identified a microwave contribution that can be emitted directly from the acceleration region, and in those cases the corresponding spectral component was also narrowband, in contrast to the broadband contribution normally seen from the trapped looptop component. Based on this prominent distinction in properties of these two components, in this paper we undertake a systematic search for and analysis of these narrowband GS bursts.

GS microwave spectra typically display a distinct spectral peak with the low-frequency and high-frequency tails described roughly by a power-law in frequency.  GS microwave bursts are broadband continuum emission from the combined thermal and non-thermal population of electrons, hence a broad bandwidth can arise when either or both of the low-frequency or high-frequency slopes are somewhat flat. The flatness of the high-frequency slope is seen when the power-law energy spectrum of the fast electrons is relatively hard, while the flatness of the low-frequency slope is mainly due to the source inhomogeneity. Therefore, acceleration of fast electrons with a soft (steep) energy spectrum into a reasonably uniform source would result in the generation of a narrowband GS burst. Clearly, instruments with a few frequency channels are intrinsically unable to provide any detailed information on a narrowband burst, or even to detect a burst that is restricted to frequencies entirely between two adjacent channels. 
Thus, it is not surprising that the studies of the narrowband bursts are limited to a few cases mentioned above and \citet{2006A&A...457..693G}, who observed a relatively narrowband GS component visible for a few seconds at a rise phase of a more broadband microwave burst. Therefore, it is entirely unclear how frequent such narrowband bursts are, what are their common properties, and to what extent they are different from more well-known broadband microwave bursts. Narrowband GS bursts can only be reliably detected by an instrument with high spectral resolution, like the Owens Valley Solar Array (OVSA). In this paper we study the subset of bursts seen with OVSA during 2001-2003 that showed both low- and high-frequency slopes greater than 3.

\section{OVSA database and burst selection}

Owens Valley Solar Array \citep[OVSA,][]{1984SoPh...94..413H,1994ApJ...420..903G} performed the microwave observations with the required high spectral resolution ($\Delta f/f \le 0.15$) sufficient to detect and resolve bursts with bandwidth of a few GHz. \citet{2004ApJ...605..528N} studied the statistical properties of 412 microwave bursts observed by OVSA during 2001-2002, and found that both high-frequency and low-frequency spectral power-law indices ($\delta_{hf}$ and $\delta_{lf}$, respectively) of the observed spectra have rather broad distributions in the ranges $-6 < \delta_{hf}< 0$, $0<\delta_{lf}<6$, which implies that in many bursts either or both low-frequency or high-frequency slopes are rather steep. The bursts with steep slopes on both sides represent the subclass of the narrowband bursts we focus on.

To perform a statistical study of this burst subclass,  we automatically identified bursts displaying sufficiently steep slopes, $|\delta_{lf,hf}| \ge 3$ at each side of the spectrum, at least for one time frame. Then, we manually discarded all events that were found to have been selected due to statistical outliers. The selection of steep spectra with $\delta_{lf}> 3$ is motivated by the fact that such spectra cannot be produced by GS self-absorption \citep[limited to $\delta_{lf} \approx 2.9$, c.f.][]{Dulk_Marsh_1982} and so require some additional process to be involved, e.g. Razin suppression or/and free-free absorption \citep{2007ApJ...666.1256B}, both of which indicate presence of a dense thermal plasma.

\begin{figure*}
\epsscale{0.95} \plotone{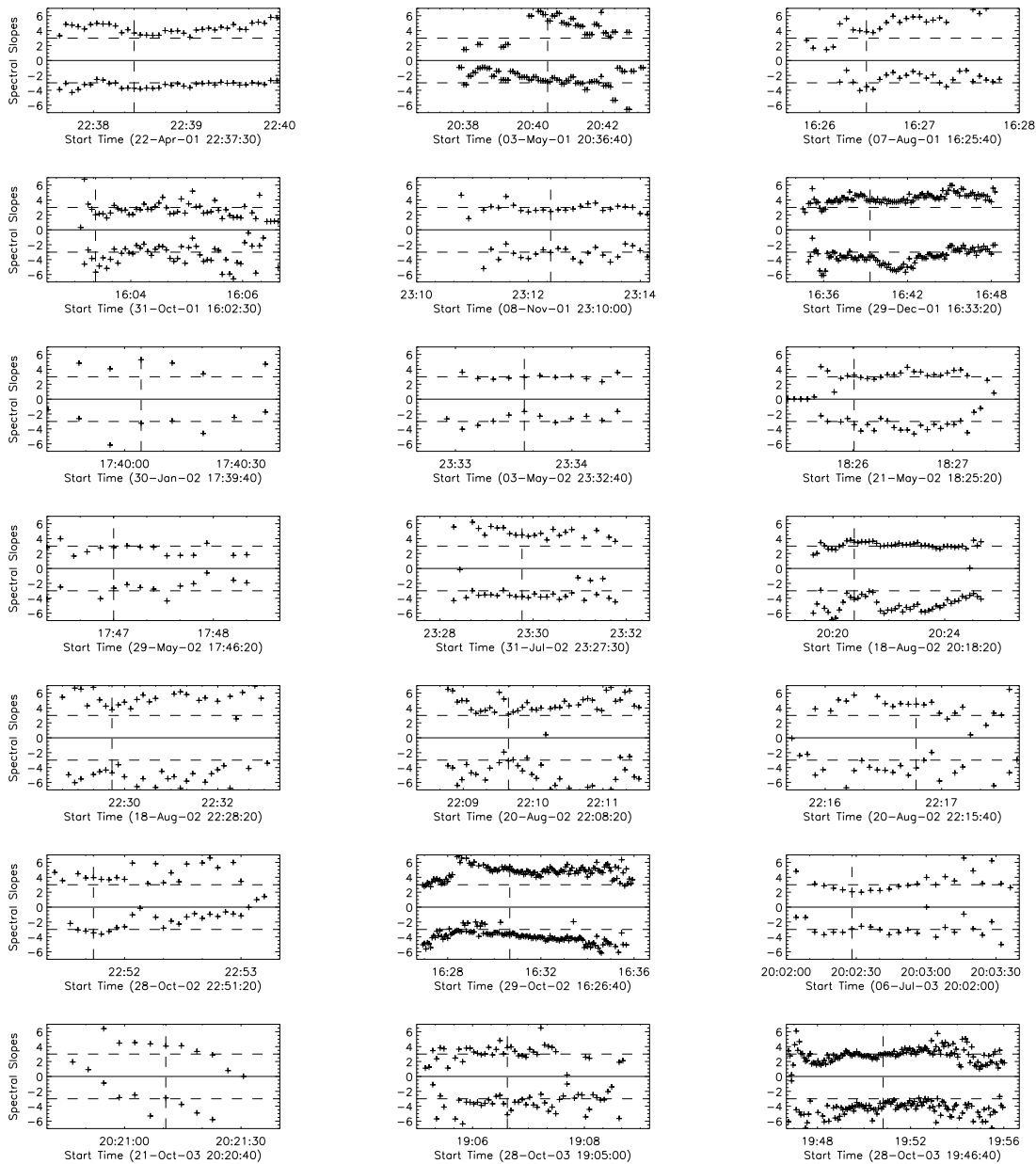}
\figcaption{\small Evolution of low-frequency (positive, symbols) and high-frequency (negative, symbols) spectral slopes, $F_f\propto f^{\delta_{lf,~hf}}$
for the 21  narrowband microwave bursts determined from the sequential fit of the instantaneous OVSA spectra to a so-called generic function. The dashed horizontal lines correspond to $\delta_{lf}=3$ and $\delta_{hf} =-3$ used to automatically classify a burst as a narrowband one. Dashed vertical lines indicate time of the absolute peak flux.
\label{slopes_narrow}}
\end{figure*}

We found that about 10$\%$ of all bursts observed by OVSA during 2001-2003 (more than 40 bursts) fall into this subclass, displaying steep slopes at both low and high frequencies during at least part of the burst duration. We quantitatively analyze a subset of 21 well isolated bursts with clear rise and decay phases allowing quantitative analysis---other bursts had either too short duration, or strong noise, or some uncorrectable instrumental effects.  The dynamic spectra of these 21 bursts are shown in Figure~\ref{spectra_narrow}.  We note that some of the radio bursts are stand-alone ones, while others represent a subburst (or a precursor) of a bigger event. Quantitative information about the spectral shape in shown in Figure~\ref{slopes_narrow}, where the low- and high-frequency radio spectral indices are shown for the entire duration of each burst.  The points are 4 s apart (the time resolution of OVSA).

\section{Overview of the events }

The spectral narrowness could in principle be dependent on the geometry/position of the events, such as limb or occulted flares. However, from the imaging data presented below in Sec.~\ref{S_imaging}, it is apparent that there is no preferential position of these flares on the solar disk. Only four of them are limb flares, while others originated from various disk locations showing a reasonably uniform distribution over the disk.

\begin{figure*}
\epsscale{0.95} \plotone{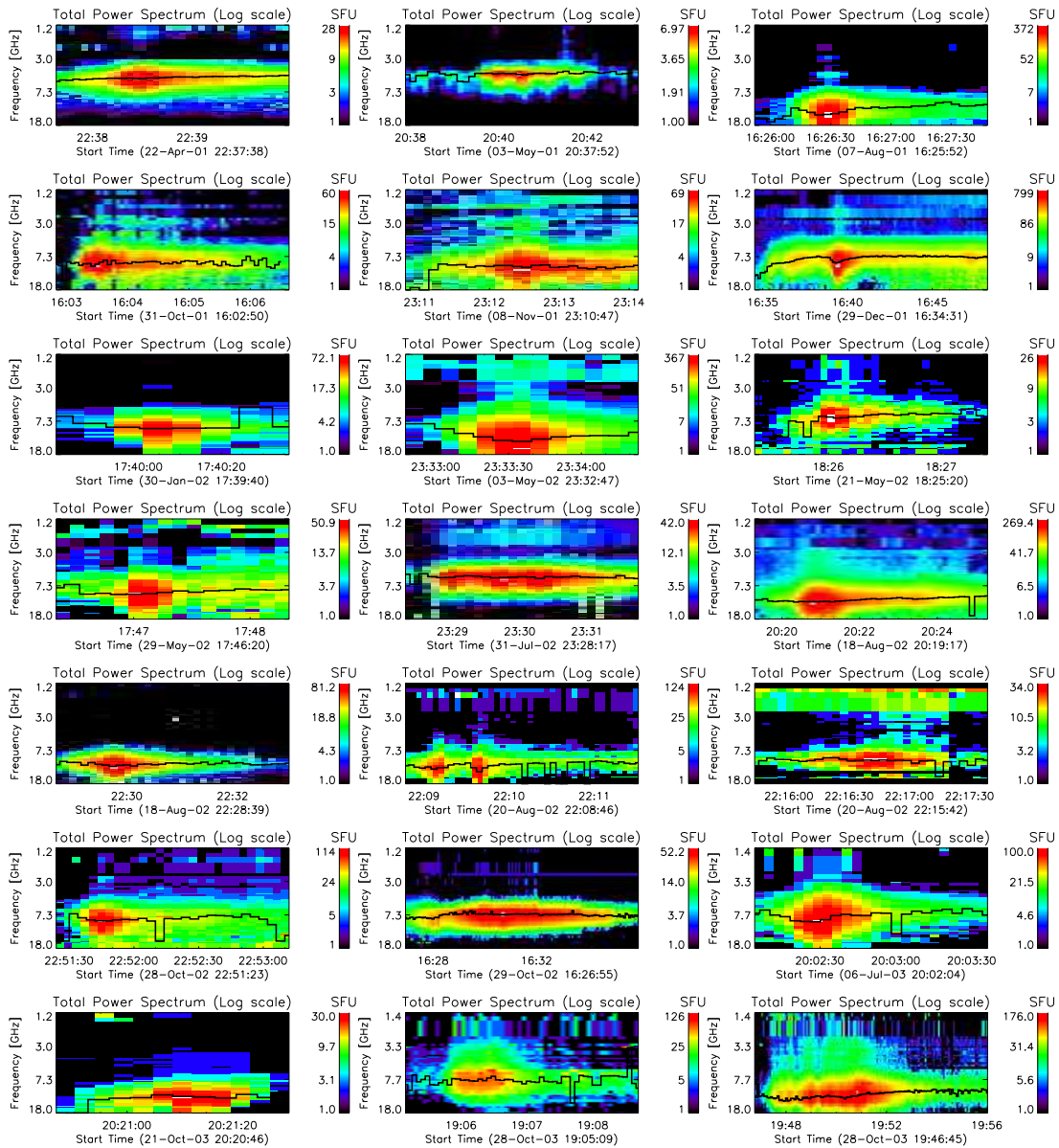} \figcaption{\small Dynamic
spectra of 21 selected narrowband microwave bursts observed with OVSA at 40
frequencies in the [1.2--18] GHz range and with 4 s time
resolution. Black curves display position of spectral peak frequencies derived from sequential fitting of the instantaneous microwave spectra.   Note that most of the dynamic spectra are only seen at a subset of available frequencies. Such bursts would be only partially observed or entirely missed with observations at single frequencies with a large spacing between them, e.g., by Nobeyama polarimeters.  \label{spectra_narrow}}
\end{figure*}

An important characteristic of microwave bursts is the spectral peak frequency, which is shown in each panel of Figure~\ref{spectra_narrow} by black curves. In most of the events, this peak frequency is almost independent of the peak flux. This suggests that the Razin-effect plays the primary role in forming the spectral shape at low frequencies \citep{2008SoPh..253...43M}, requiring a rather high thermal plasma density, typically $n_0 \sim 10^{11}$~cm$^{-3}$. However, in five cases, namely, 07-Aug-01, 29-Dec-01, 03-May-02, 20-Aug-02 22:08, and 06-Jul-03, the spectral peak frequency increases during the peak of the radio burst. As discussed in \citet{2008SoPh..253...43M}, this is indicative of the GS self-absorbtion effect dominating over the Razin-effect during these short time intervals.

The considered radio bursts are typically short lasting only a few minutes or less, with a few exceptions lasting almost 10 minutes. Most of them (18) display a single peak in the light curves, others have two or more peaks---these two spectral peaks are explicitly shown in the dynamic spectrum of the 2002-Aug-20 22:08 burst in Fig.~\ref{spectra_narrow}, while for the bursts of 2001-May-03 and 2002-Aug-18 22:28 the first temporal peaks are not shown. As mentioned earlier, in some cases the isolated narrowband bursts are followed by a stronger and more broadband burst.

\begin{figure*}
\epsscale{0.95} \plotone{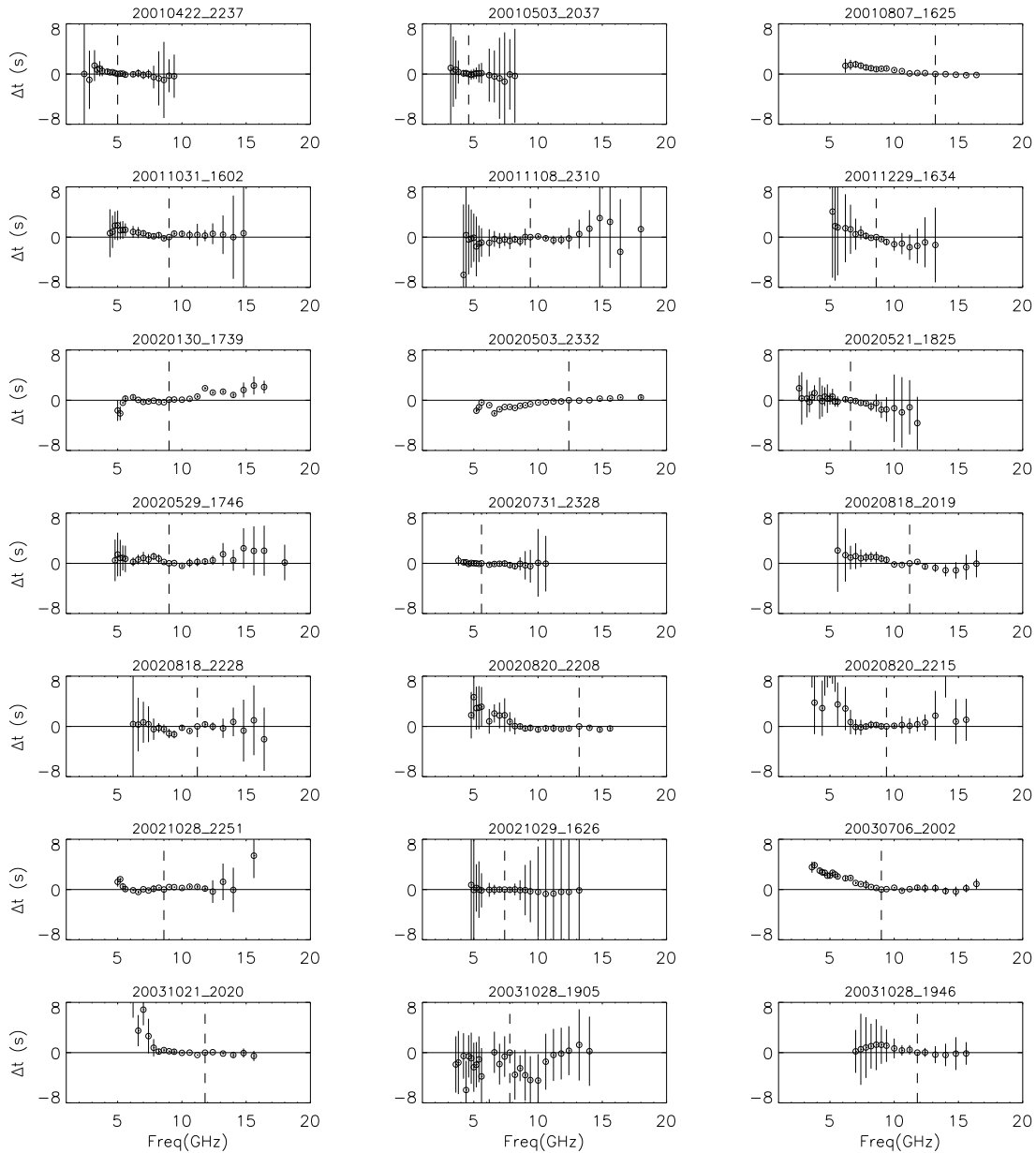}
\figcaption{\small Time delay of light curves at different frequencies vs the light curve at the frequency of the absolute peak flux for the 21  narrowband microwave bursts determined with lag cross-correlation analysis. The vertical dashed curves show positions of the absolute peak frequency.
\label{delays_narrow}}
\end{figure*}

As noted by \citet{2007ApJ...666.1256B}, if the Razin-effect is important, then the free-free absorption in the source can be important, too, when its temperature is not too high, say $T\lesssim 10$~MK. In such cases, the plasma heating, as it progresses during the course of the flare, will reduce the free-free absorption as a function of time, which will lead to a progressive delay of the low-frequency light curves relative to high-frequency light curves \citep{2007ApJ...666.1256B}. We checked this expectation in our data sample, as shown in Figure~\ref{delays_narrow}, and found that only three events convincingly show the progressive delay toward low frequencies (pronounced negative slope in the figure). Six more events suggest a similar trend but with larger error bars, and are hence not significant, while the remaining twelve events either do not show any delay, or show negative delay (positive slope in the figure). Given the high plasma density inferred from the dominance of the Razin effect in these events, the lack of prominent delay of the low-frequency light curves in most of the cases suggests that no significant local plasma heating occurs in the radio sources during the burst peak time.

\begin{figure*}
\epsscale{0.95} \plotone{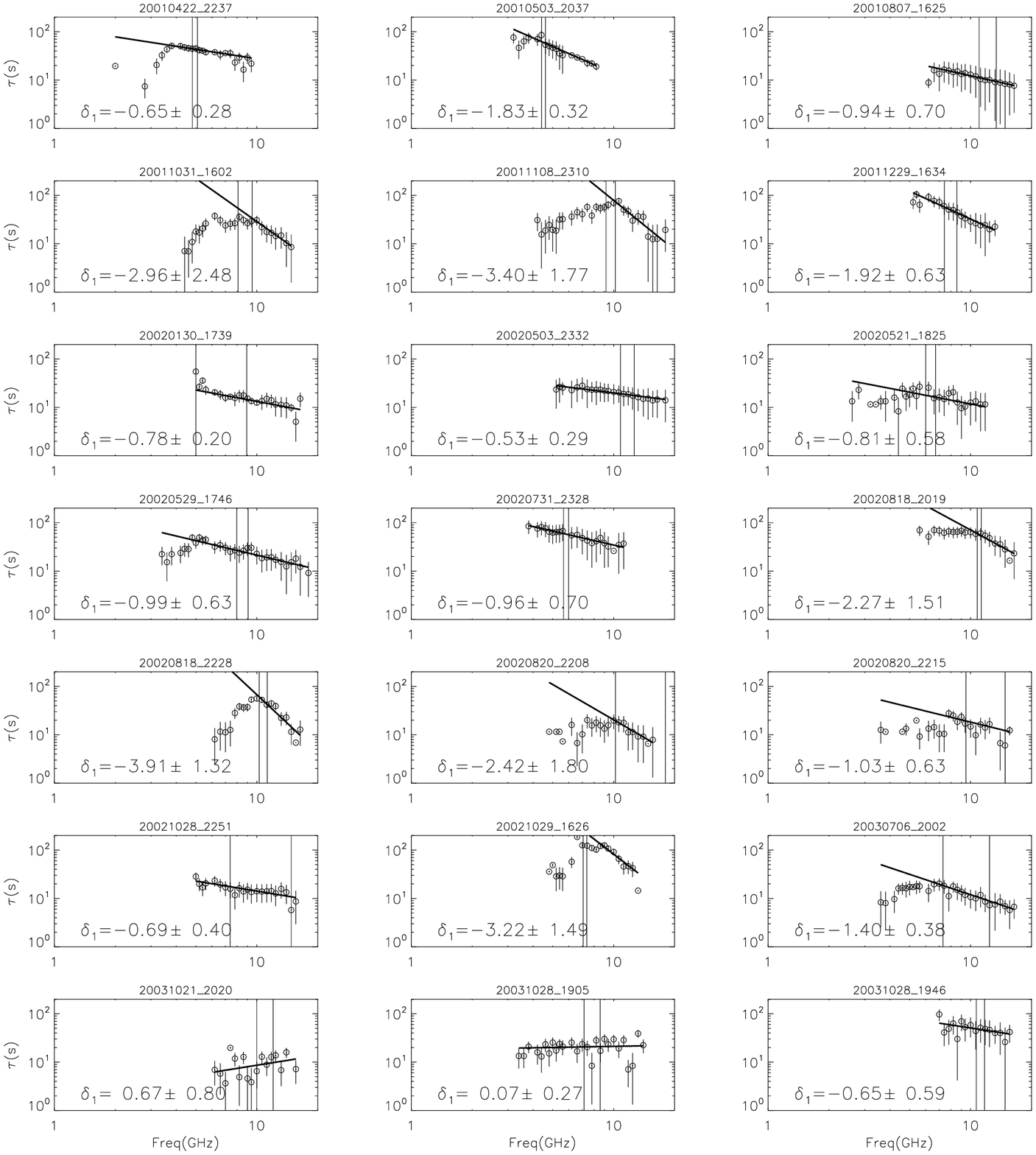}
\figcaption{\small
Exponential decay time constants (symbols) of the microwave light curves at different frequencies for the time interval immediately following the peak time for the 21 selected narrowband microwave bursts and corresponding power-law fits (solid lines), $\tau(f) \propto f^{\delta_1}$, of high-frequency parts of the decays.
Note, that in almost all cases the decay constants display a statistically significant frequency dependence and it decreases with frequency in most of the cases.
Two vertical solid lines in each panel indicate the range in which the spectral peak frequency during the time interval selected to analyze the decay.
\label{decay_1_narrow}}
\end{figure*}

\begin{figure*}
\epsscale{0.95} \plotone{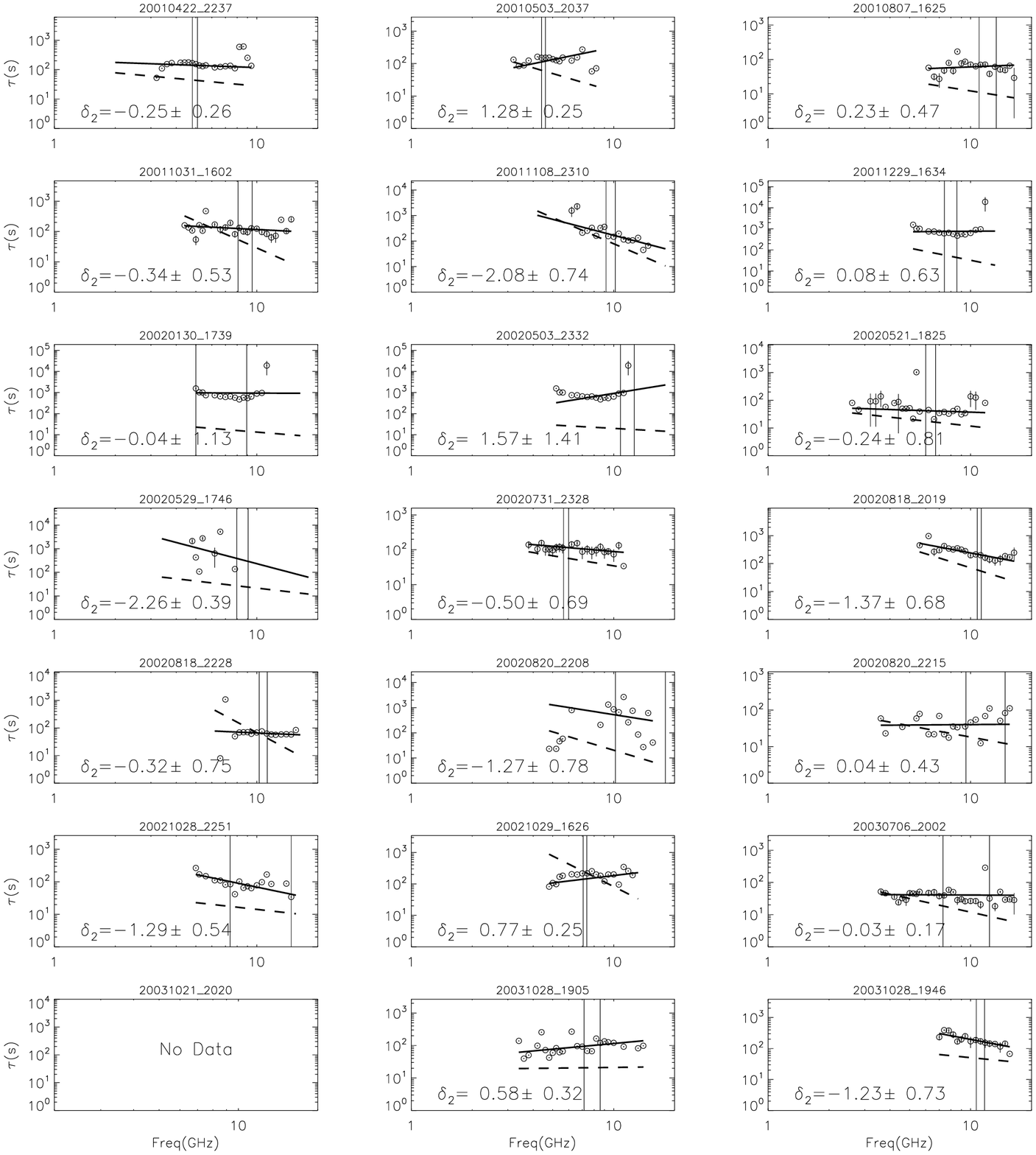}
\figcaption{\small
Exponential decay time constants (symbols) of the microwave light curves at different frequencies at late decay phase for the 21  narrowband microwave bursts and corresponding power-law fits (solid lines), $\tau(f) \propto f^{\delta_2}$,  of the decays. Dashed curves display fits from Figure~\ref{decay_1_narrow}.
Two vertical solid lines in each panel indicate the range in which the spectral peak frequency during the time interval selected to analyze the decay.
\label{decay_2_narrow}}
\end{figure*}

While the steepness of the low-frequency slope of the radio spectrum requires the Razin-effect and can be further affected by free-free absorption in the source or along the line of sight,
the optically-thin, high-frequency slope of the spectrum is related only to the electron properties. Electrons with soft spectral slope ($\delta \gg1$) or  alternatively a relatively small value of the high-energy cut-off ($E_{\max} \simeq$ a few $\times100$~keV) can produce a narrow GS spectrum. In the latter case most of the optically thin GS emission is produced by the electrons with $E\sim E_{\max}$.  However, if the microwave emission at all optically thin frequencies is produced by the electrons with $E\sim E_{\max}$, then the light curves at all these frequencies must be proportional to each other, with no delay between them and with identical, frequency-independent decay constants. Although Fig.~\ref{delays_narrow} offers support for no delay, Figures~\ref{decay_1_narrow} and  \ref{decay_2_narrow} show clearly that the decay constants are generally not independent of frequency, but rather tend to show faster decay at higher frequencies.  This means that, despite the narrow spectra, emission at various frequencies does preferentially originate from electrons with different energies, and so the study of light curves at different frequencies will contain information on the electron evolution at different energies, which we discuss in more detail later.

In addition, taking into account that the narrow bandwidth of these bursts is strong evidence of a reasonably homogeneous source, we explicitly fit the observed sequence of the spectra by the gyrosynchrotron source function of a uniform source to derive the source parameter evolution; see Sec.~\ref{S_mw_fit}.

\section{Radio and X-ray imaging}
\label{S_imaging}

\subsection{X-ray imaging and spectroscopy}

\begin{figure}\centering
\includegraphics[width=0.48\columnwidth]{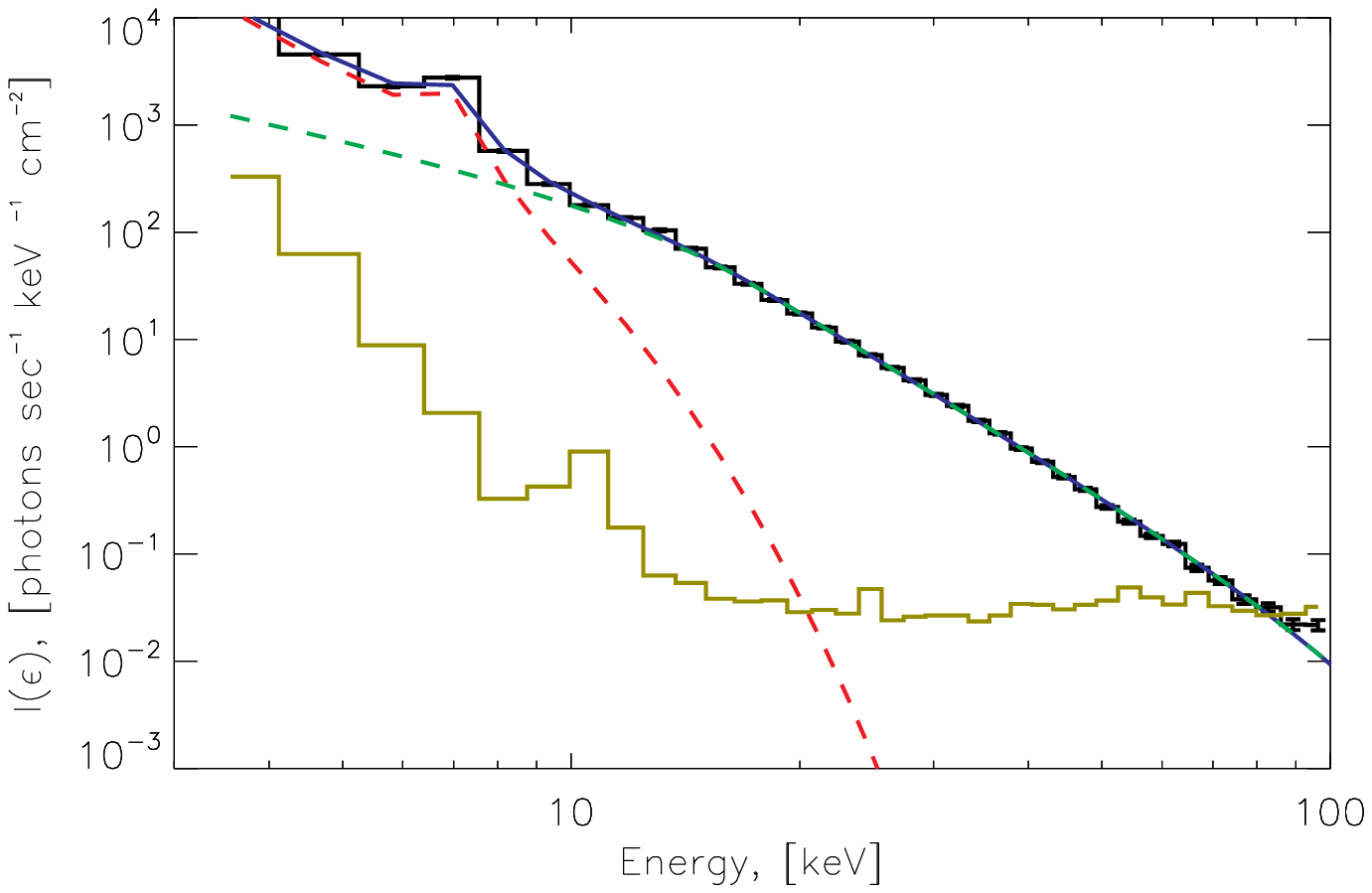}
\includegraphics[width=0.48\columnwidth]{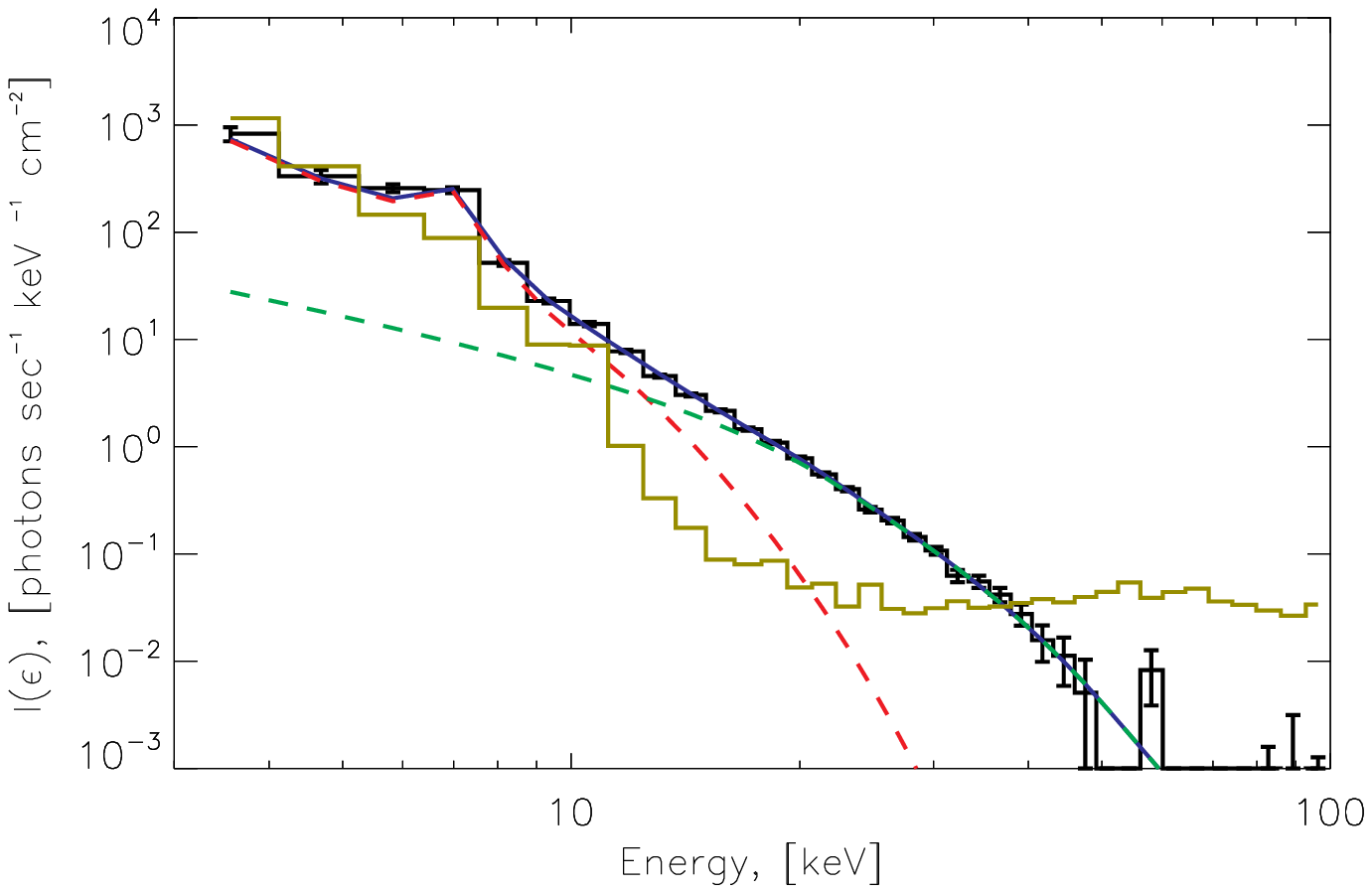}\\
\includegraphics[width=0.48\columnwidth]{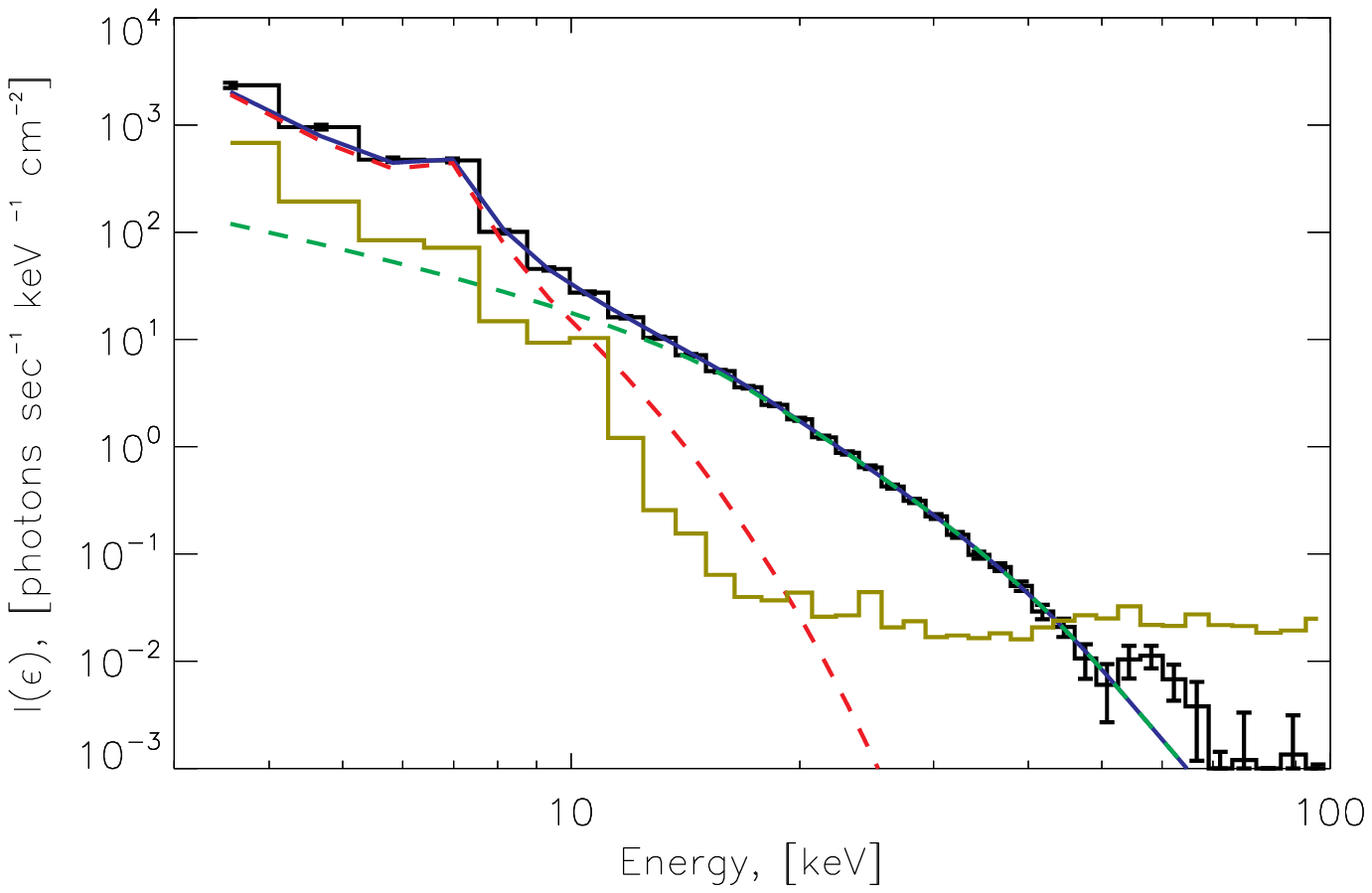}
\includegraphics[width=0.48\columnwidth]{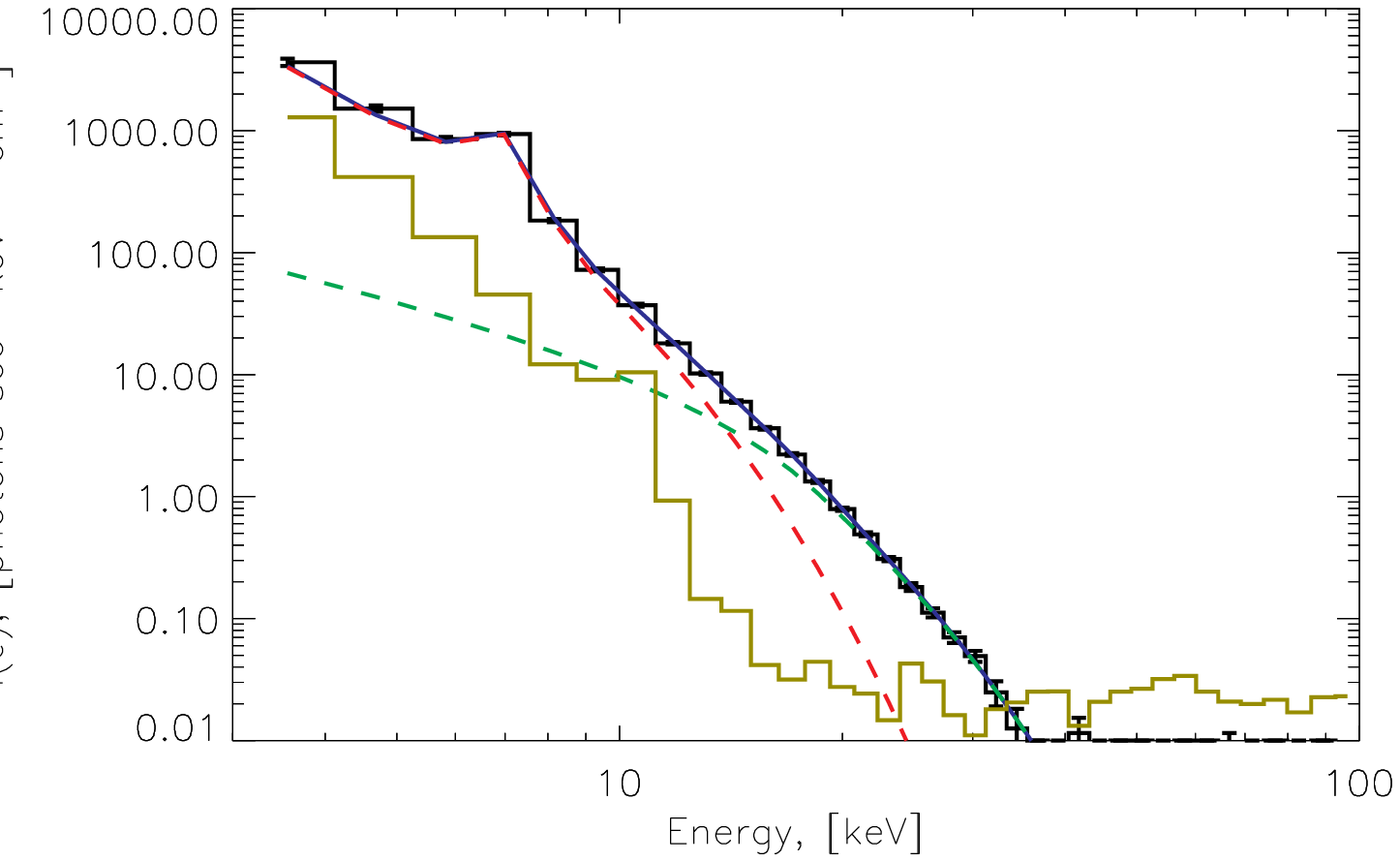}
\caption{\label{fig:Xspectr} \rhessi\  X-ray spectra fitted above 6 keV for four events (clockwise from the top left 2002-May-03 23:33:08 UT, 2002-May-21 18:25:32 UT,
2002-May-29 17:46:40 UT, 2002-July-31 23:28:52 UT). The spectrum is accumulated
over the HXR peak of the events. Photon spectrum (black histogram), iso-thermal component (red dashed),
non-thermal component assuming thick-target model (green), the total thermal plus non-thermal model (blue). }
\end{figure}

A total of 12 of our selected narrowband bursts are also available for detailed study with \rhessi\ \citep{2002SoPh..210....3L}. A spatially integrated hard X-ray (HXR) spectrum accumulated over the impulsive phase of the flare is shown in Figure \ref{fig:Xspectr} for four events. The electron properties were deduced for each flare spectrum \citep[see, e.g. review by][]{2011SSRv..159..301K} by fitting count rates with a thermal plus non-thermal thick-target model. Albedo correction \citep{2006A&A...446.1157K} was applied for all on-disk flares assuming isotropic X-ray emission. Overall, as can be seen from the spectra (Figure \ref{fig:Xspectr}), the flares show a rather strong HXR thermal component and a rather soft (large spectral index) non-thermal component.

The hard X-ray emission associated with narrowband microwave bursts is generally not very strong and is observed by \rhessi\ up to a few tens to about 100 keV. The flares where HXR data are available, in general show rather steep (soft) spectra, with the electron flux spectral index larger than $\sim 5$; see Table~\ref{distrib_table_x} for details. Such flares are sometimes associated with high density coronal sources \citep[e.g.][]{2004ApJ...603L.117V, 2007ApJ...666.1256B, 2008ApJ...673..576X, 2011ApJ...730L..22K, 2012A&A...543A..53G, 2014ApJ...787...86J}. However, for the analyzed flare set, the emission measures and the plasma densities inferred from the thermal X-ray sources are not particularly large, as seen from Table~\ref{distrib_table_x}.

\begin{figure*}\centering
\includegraphics[width=0.3\textwidth]{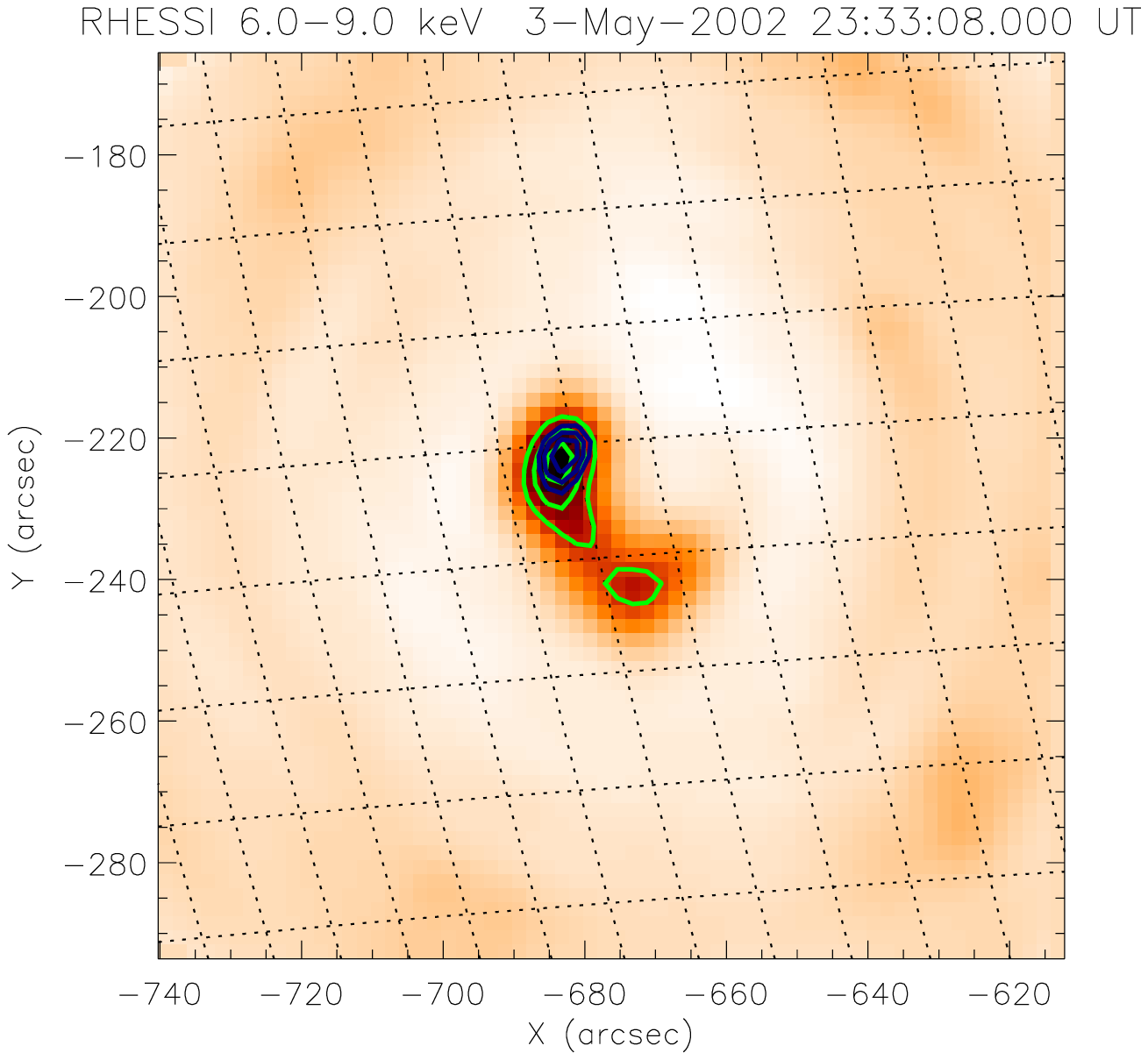}
\includegraphics[width=0.3\textwidth]{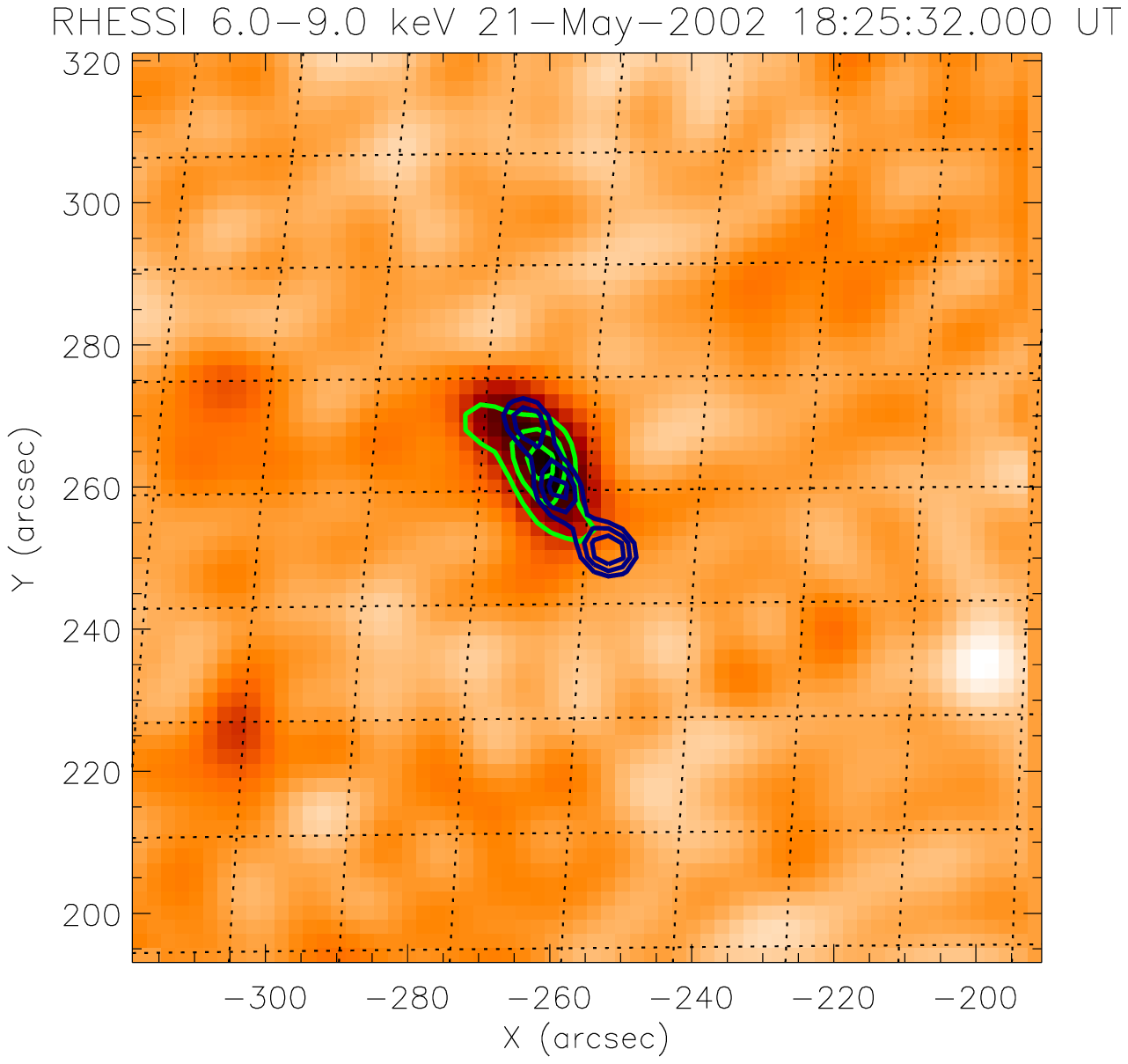}
\includegraphics[width=0.3\textwidth]{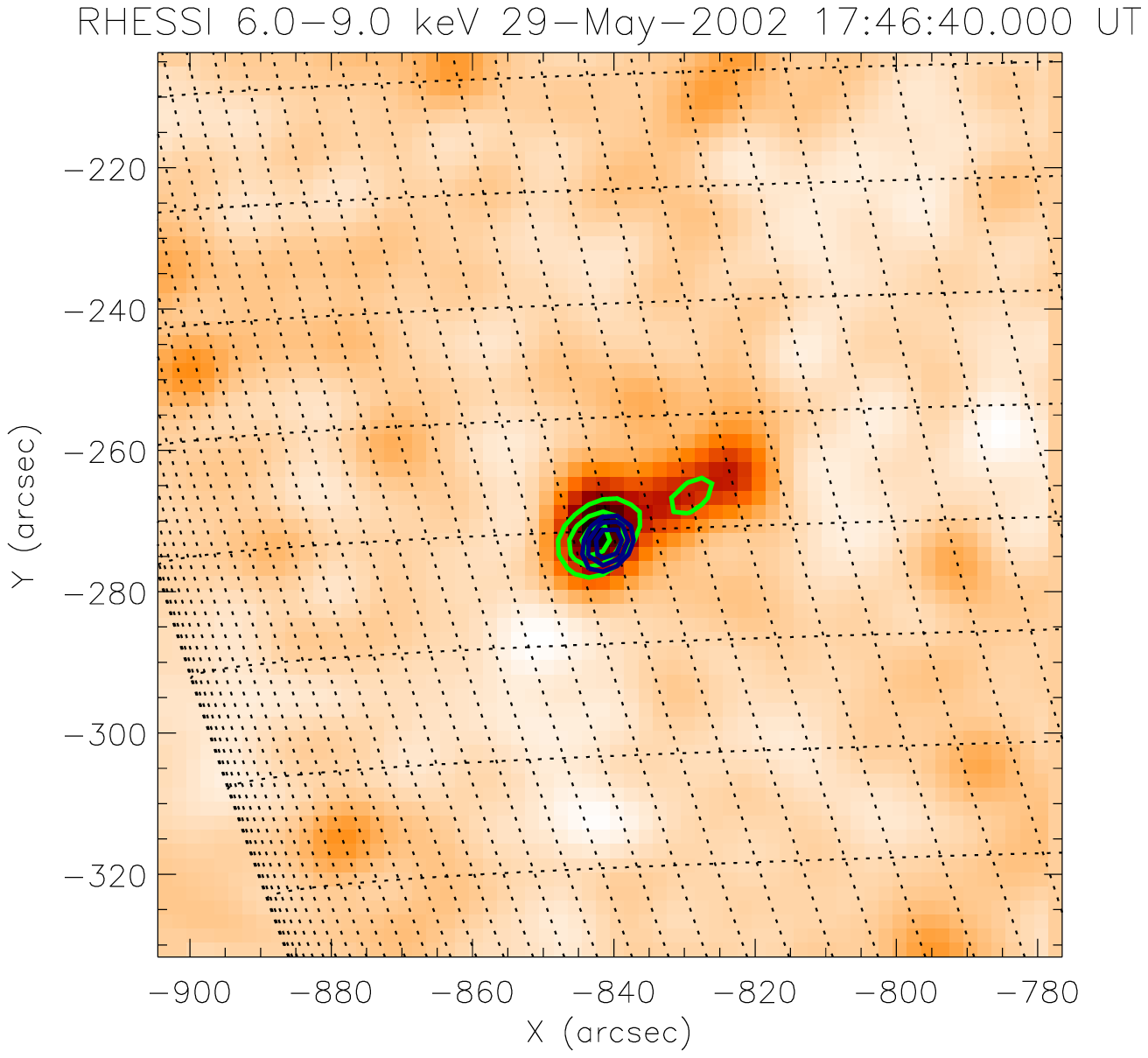}\\
\includegraphics[width=0.3\textwidth]{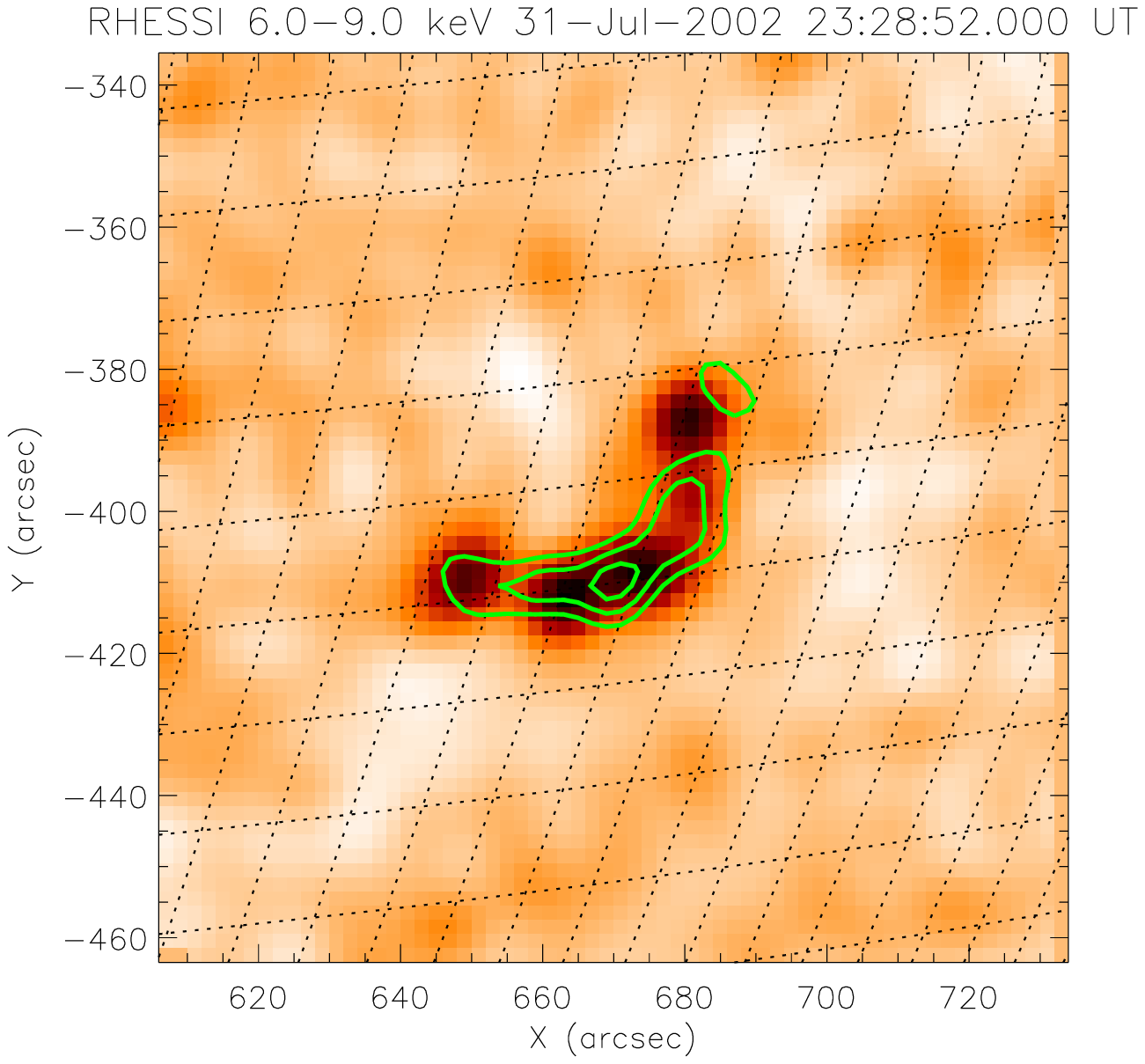}
\includegraphics[width=0.3\textwidth]{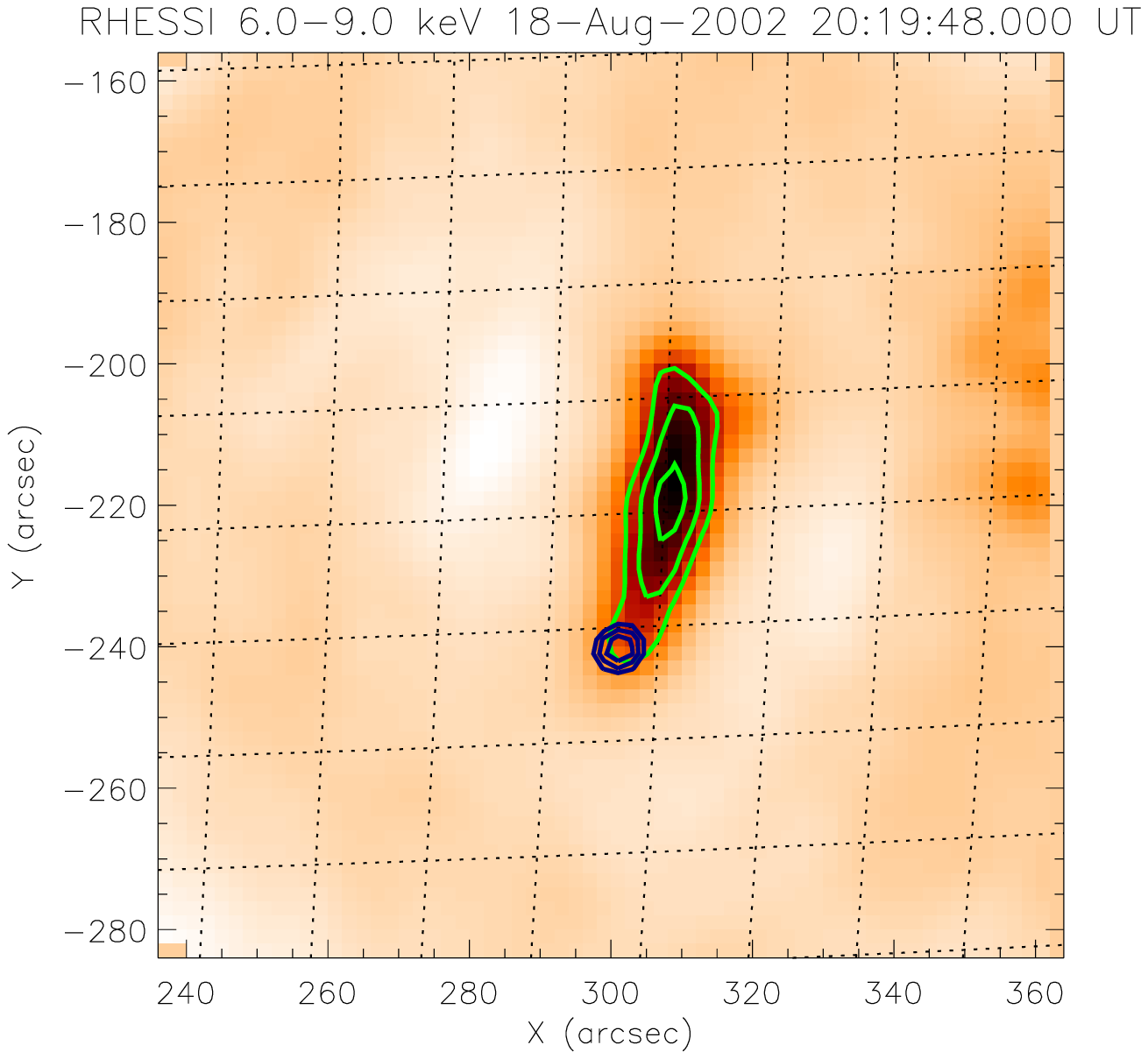}
\includegraphics[width=0.3\textwidth]{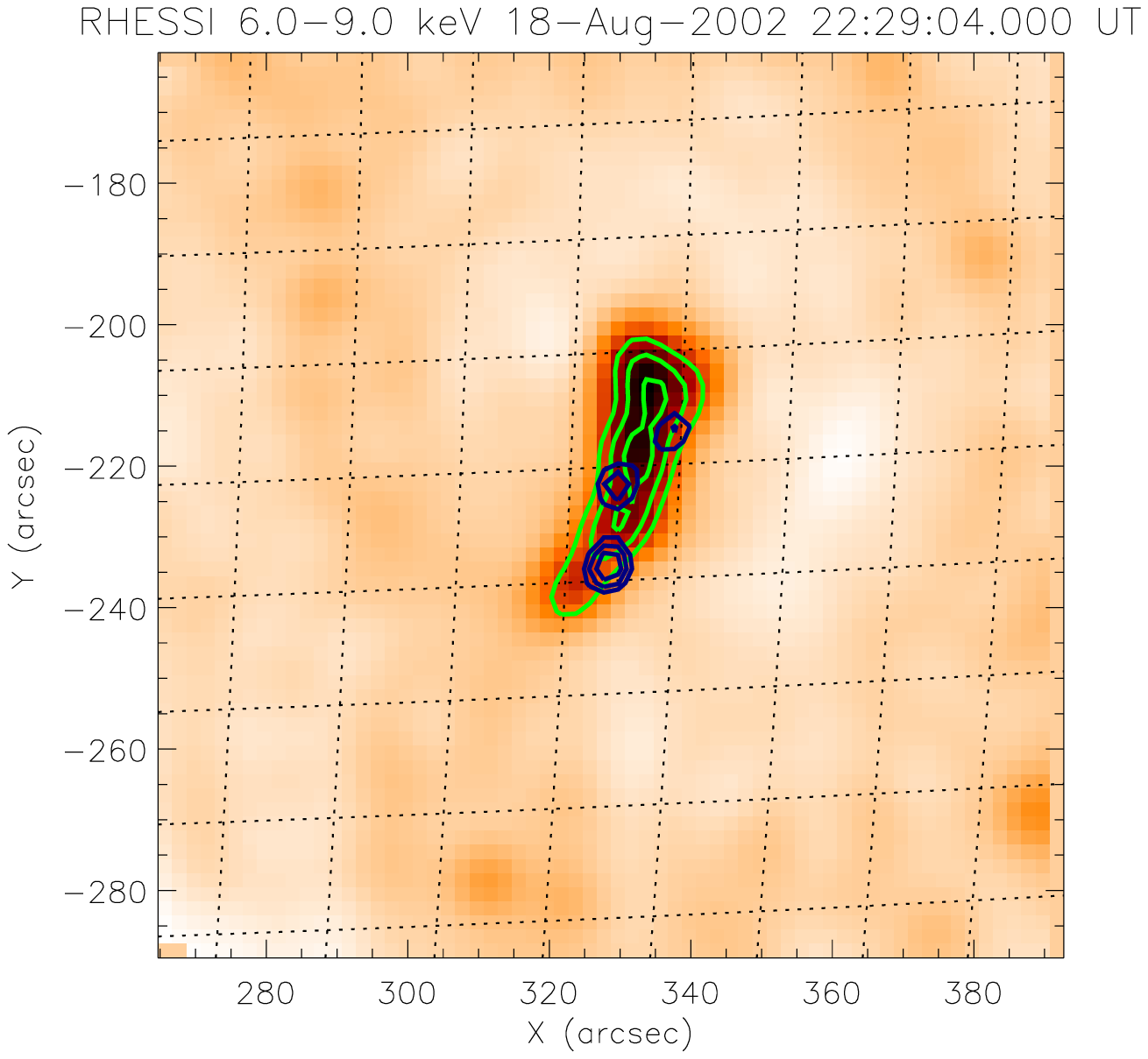}\\
\includegraphics[width=0.3\textwidth]{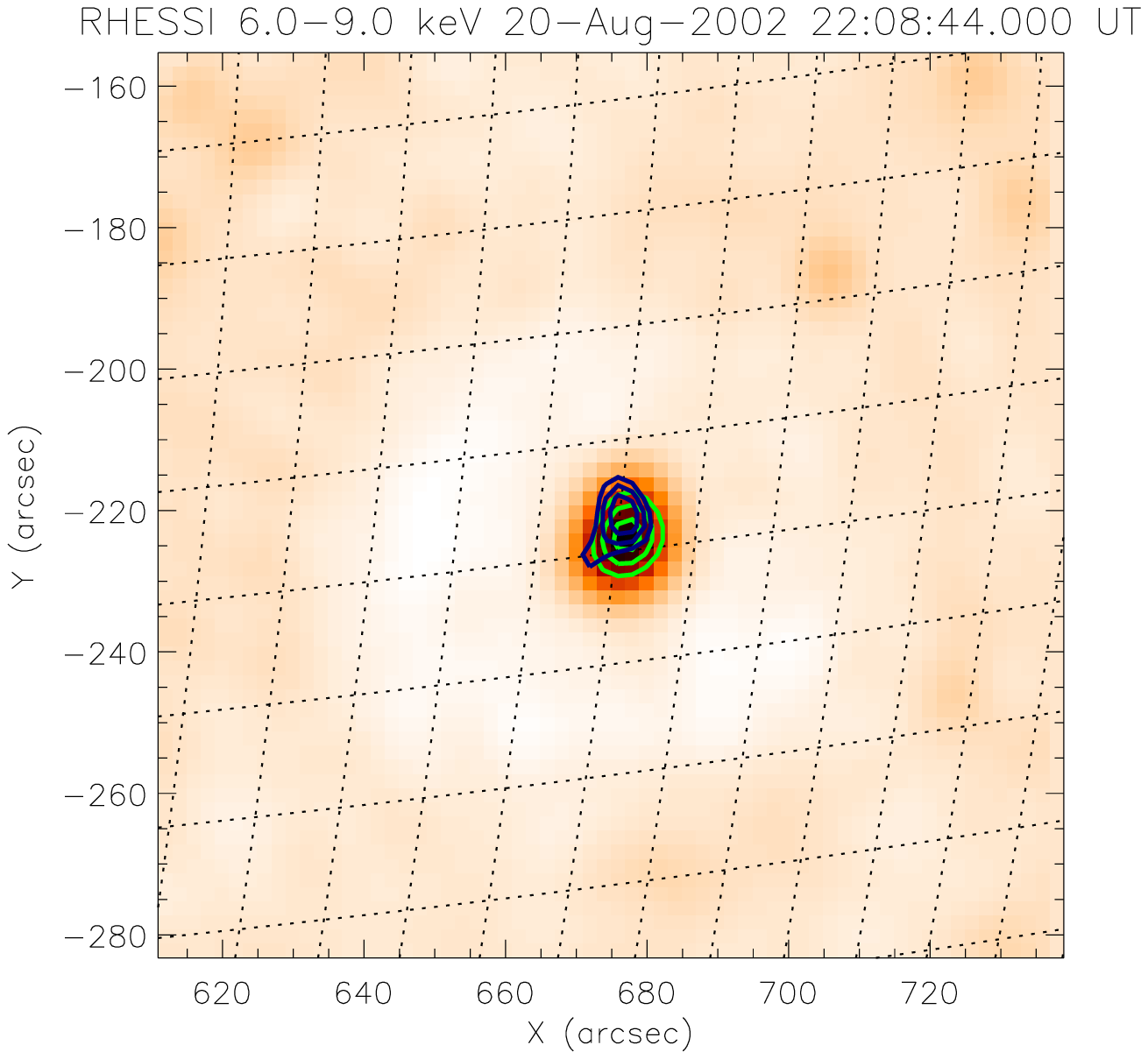}
\includegraphics[width=0.3\textwidth]{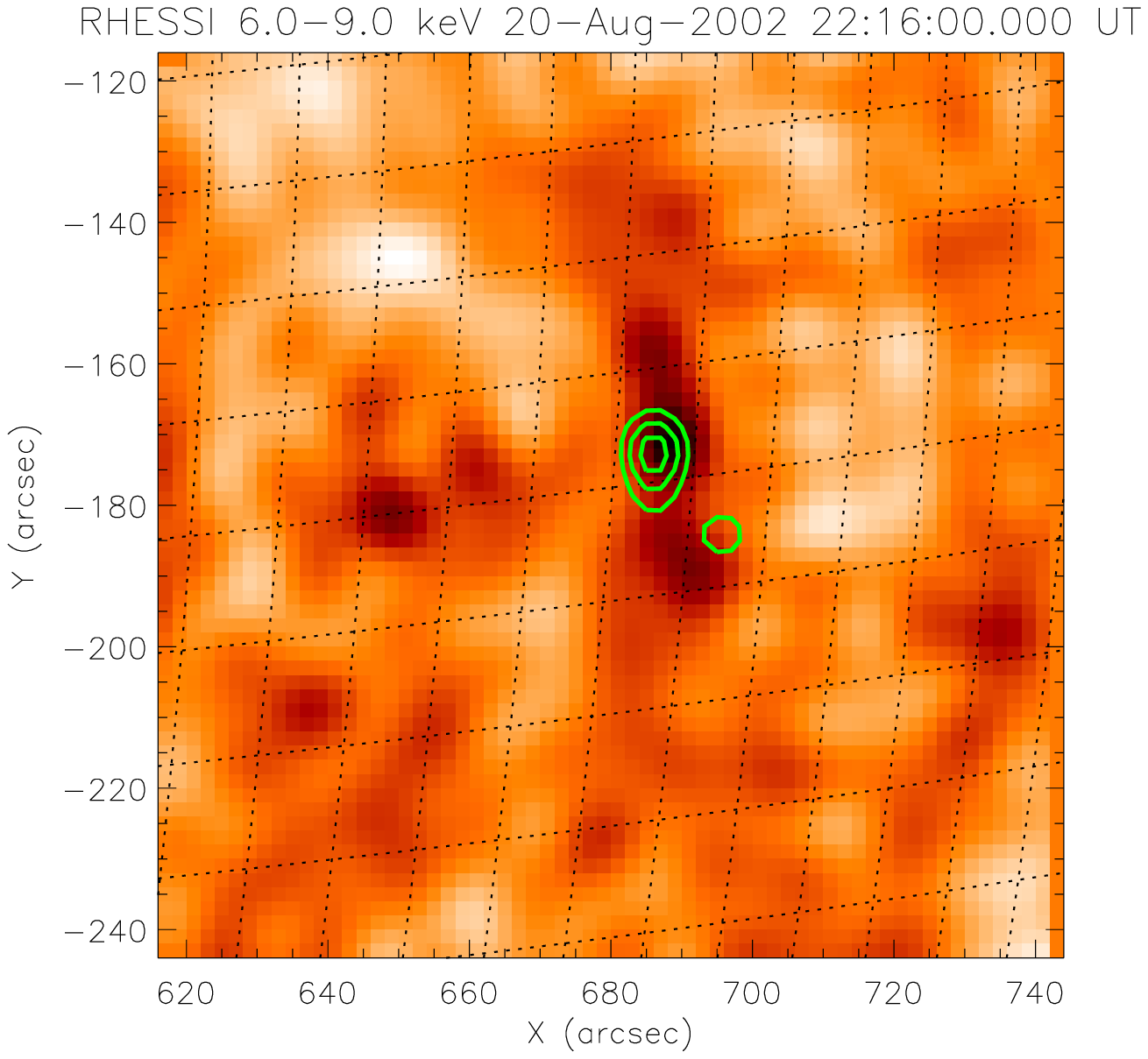}
\includegraphics[width=0.3\textwidth]{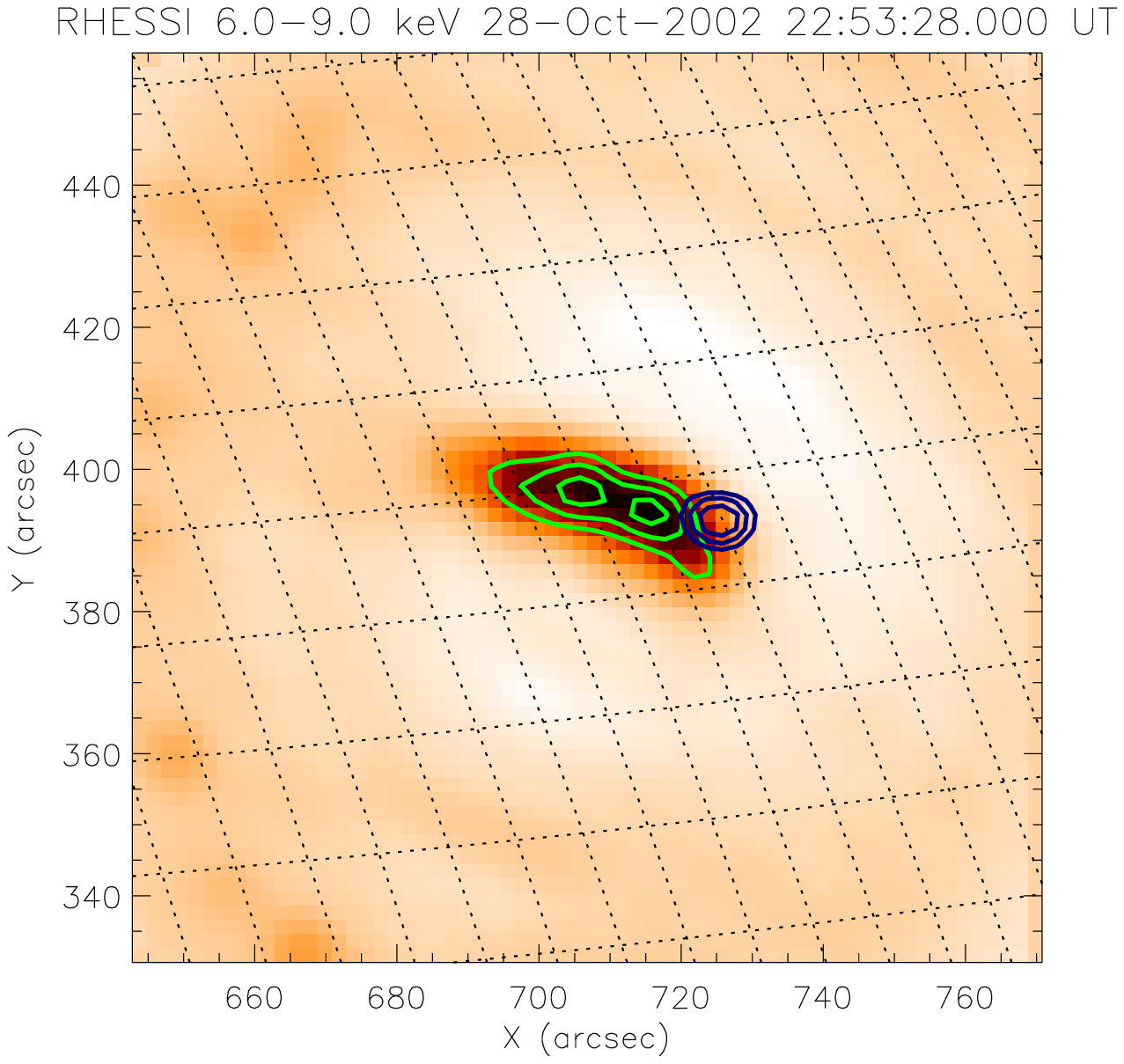}\\
\includegraphics[width=0.3\textwidth]{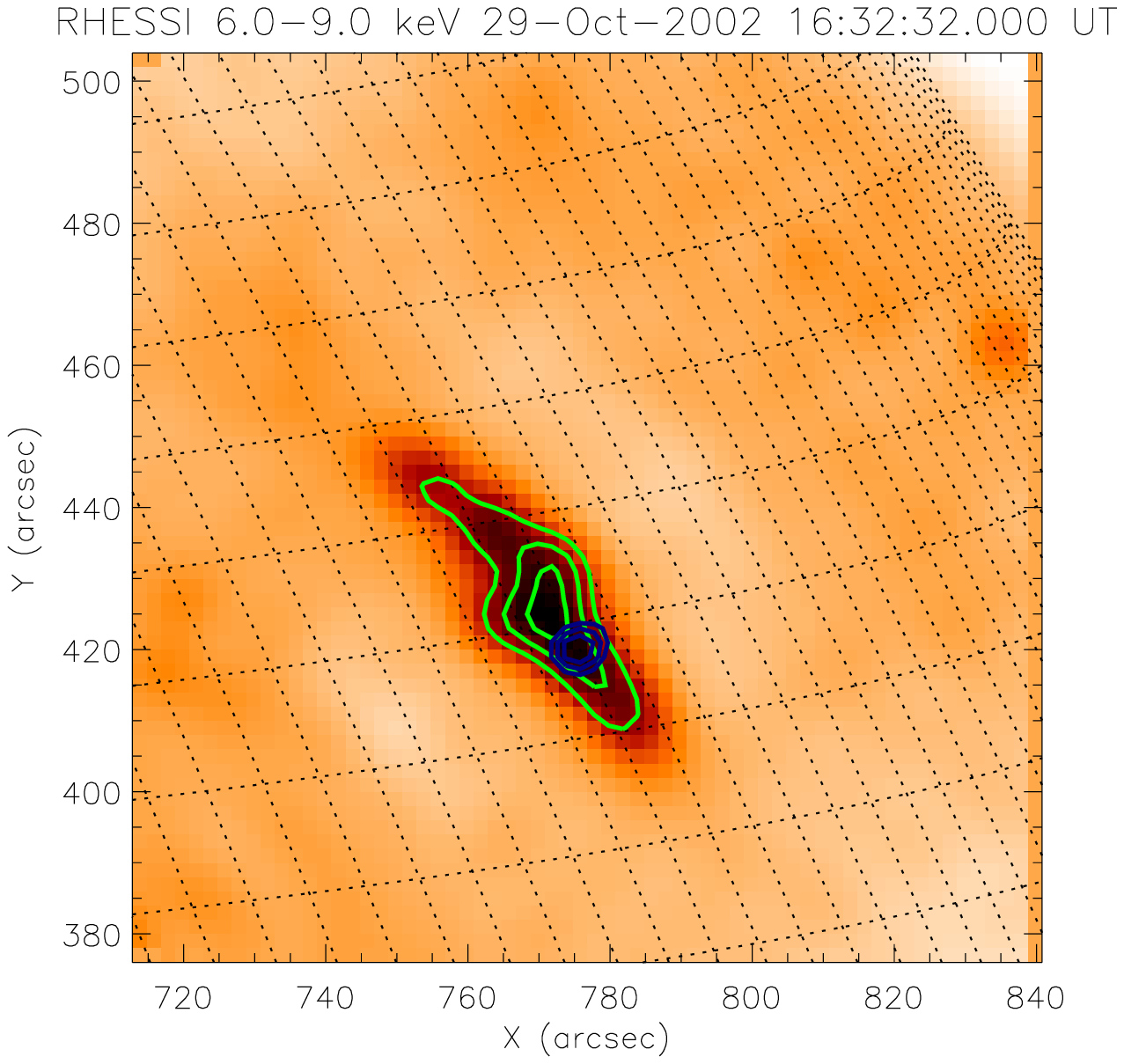}
\includegraphics[width=0.3\textwidth]{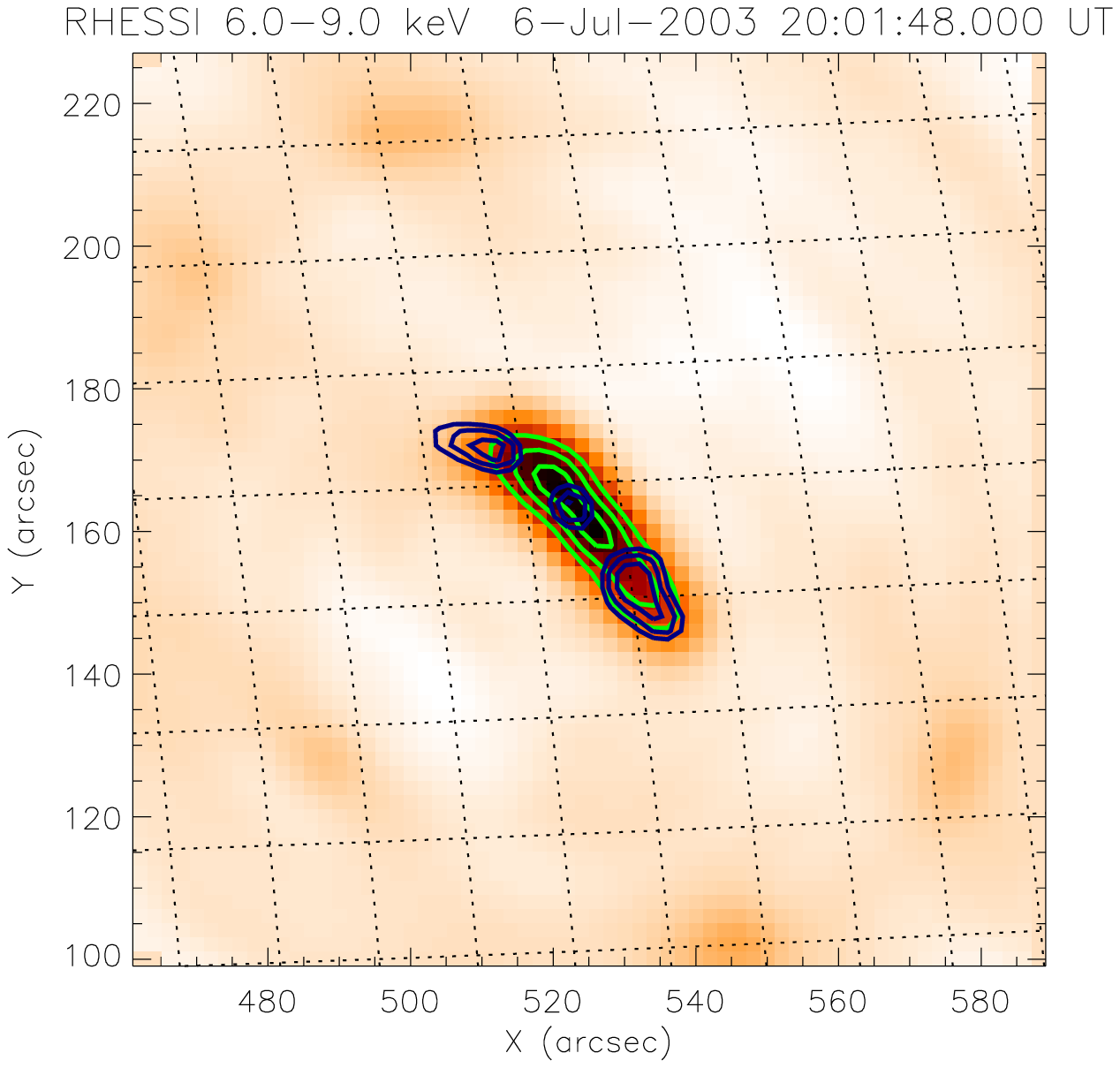}
\includegraphics[width=0.3\textwidth]{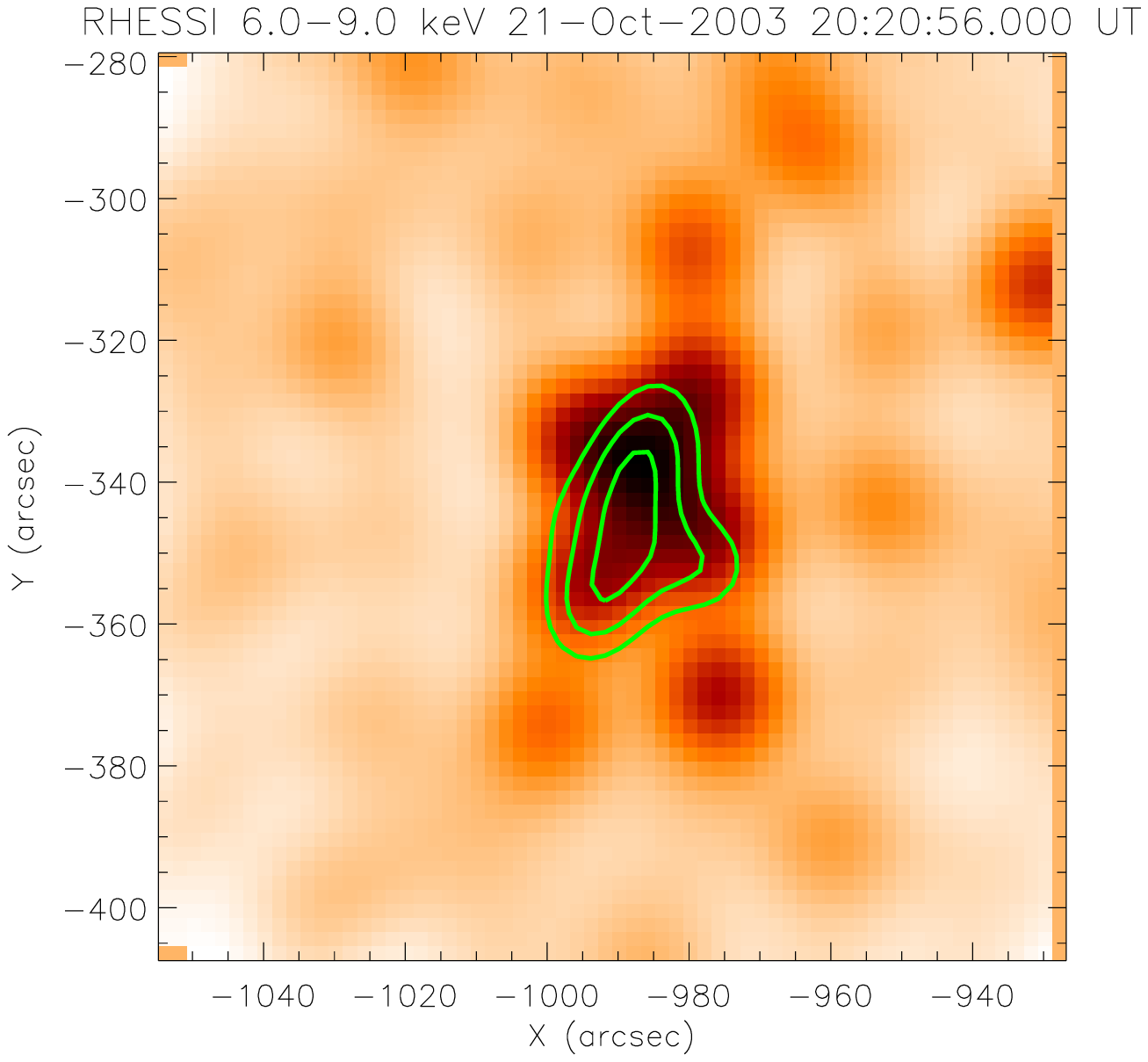}\\
\caption{\label{fig:images} \rhessi\ images (6--9~keV) and the contours at 9--15~keV (green) and 25--50~keV (blue) for 12 events.}
\end{figure*}

The HXR morphology of the narrowband bursts is likewise diverse. Out of 12-flares where HXR data allow imaging, the majority of the events show rather complicated structure.  At low energies, the emission is elongated---presumably aligned with the flaring magnetic structure. HXR emission above $\sim 25$~keV is sometimes
observed and can be imaged (Figure \ref{fig:images}) and likely to come from chromospheric footpoints, as often observed by \rhessi \citep[e.g.][]{2011SSRv..159..107H}. The events selected have one or more footpoints at higher energies. However, the number of footpoints differ from event to event (and some of the deka-keV sources can, in fact, be coronal rather than chromospheric). Only one flare 03-May-2002 shows a structure clearly consistent with a single loop morphology. In addition, the two flares on 20-Aug-2002 are compact so the structure of the flaring regions is unresolved by \rhessi. The 20-Oct-2003 event is also one flare observed high above the limb.

\subsection{OVSA imaging}

Imaging of OVSA data has recently become more accessible due to the creation of a newly-calibrated OVSA legacy data base\footnote{Currently available from Jan 2000 to Aug 2003, with recalibration for later data available soon.} and updated, more-user-friendly OVSA-imaging software \citep{2014AAS...22421845N}, which greatly facilitates imaging and image-comparison using Interactive Data Language (IDL) routines publicly available through the community contributed and maintained Solar SoftWare (SSW)\footnote{http://www.lmsal.com/solarsoft/}  IDL code repository. To produce OVSA images for a given event requires that a valid reference calibration record from long-integration measurements on cosmic sources, made a few times per year, be available for insertion into the OVSA daily raw data file.  This allows daily phase calibrations to be applied, from which an IDL \texttt{sav}-file for a given solar observation segment is created (with a typical duration of 2~hours). When a radio burst is identified in the given data record, an appropriate background is identified and subtracted, bad frequencies are flagged, the total power (TP) spectra are sequentially fit with a so-called `generic' function \citep{1989SoPh..120..351S} using the built-in functionality of the \texttt{OVSA$\_$Explorer} application, and the data are saved in the forms needed to study the calibrated TP dynamic spectra and create images. The images are generated from proprietary-formatted \texttt{uv}-files by a widget-based program, \texttt{wimagr}, also available from the OVSA tree of the SSW package.  This program has the ability to import other context images for direct comparison and overlay with the OVSA images. Various kinds of frequency and/or time synthesis methods are available, as well as the option to saving the results in various forms, including fits images, map structures, or postscript plots and video clips.

\begin{figure}
\begin{center}
\includegraphics[width=0.49\columnwidth,clip]{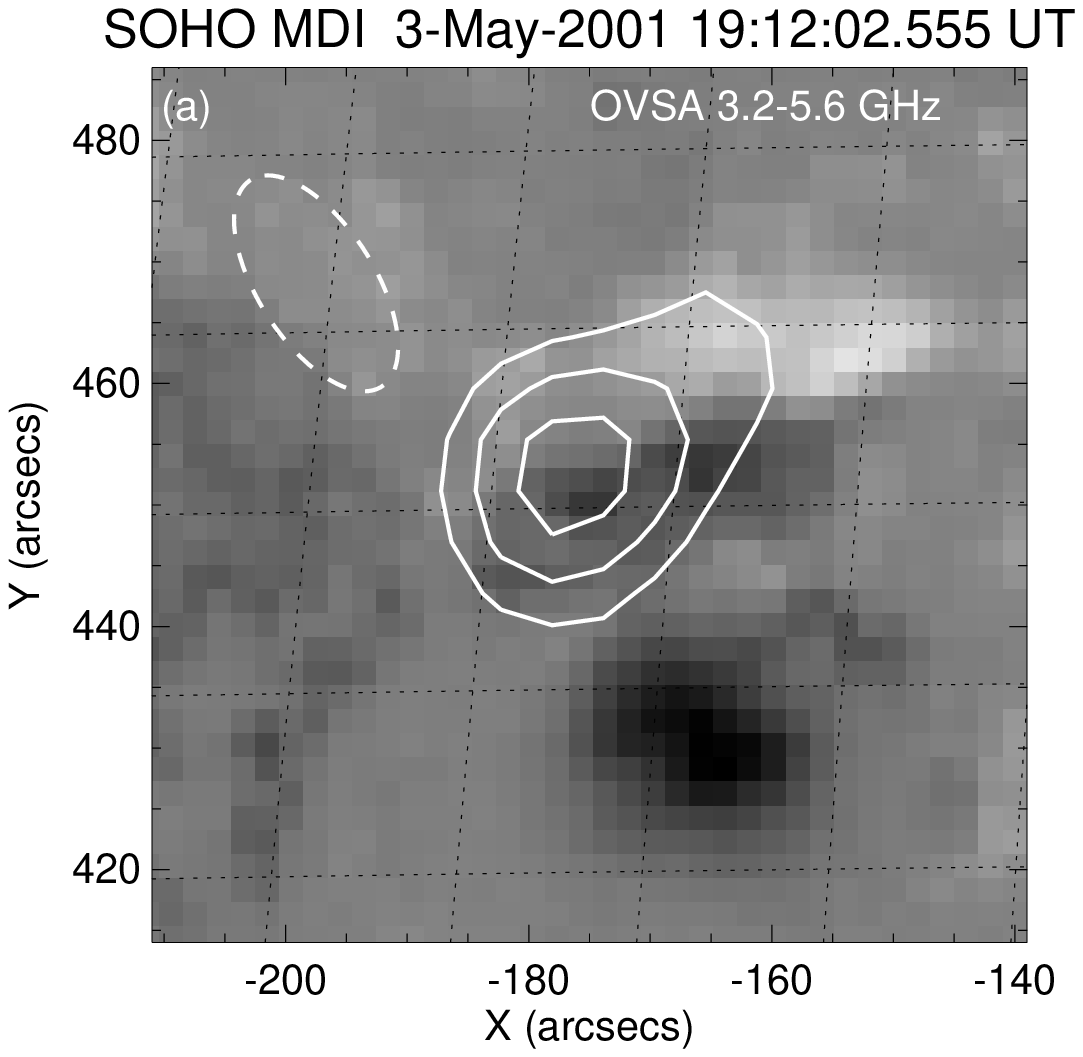}
\includegraphics[width=0.49\columnwidth,clip]{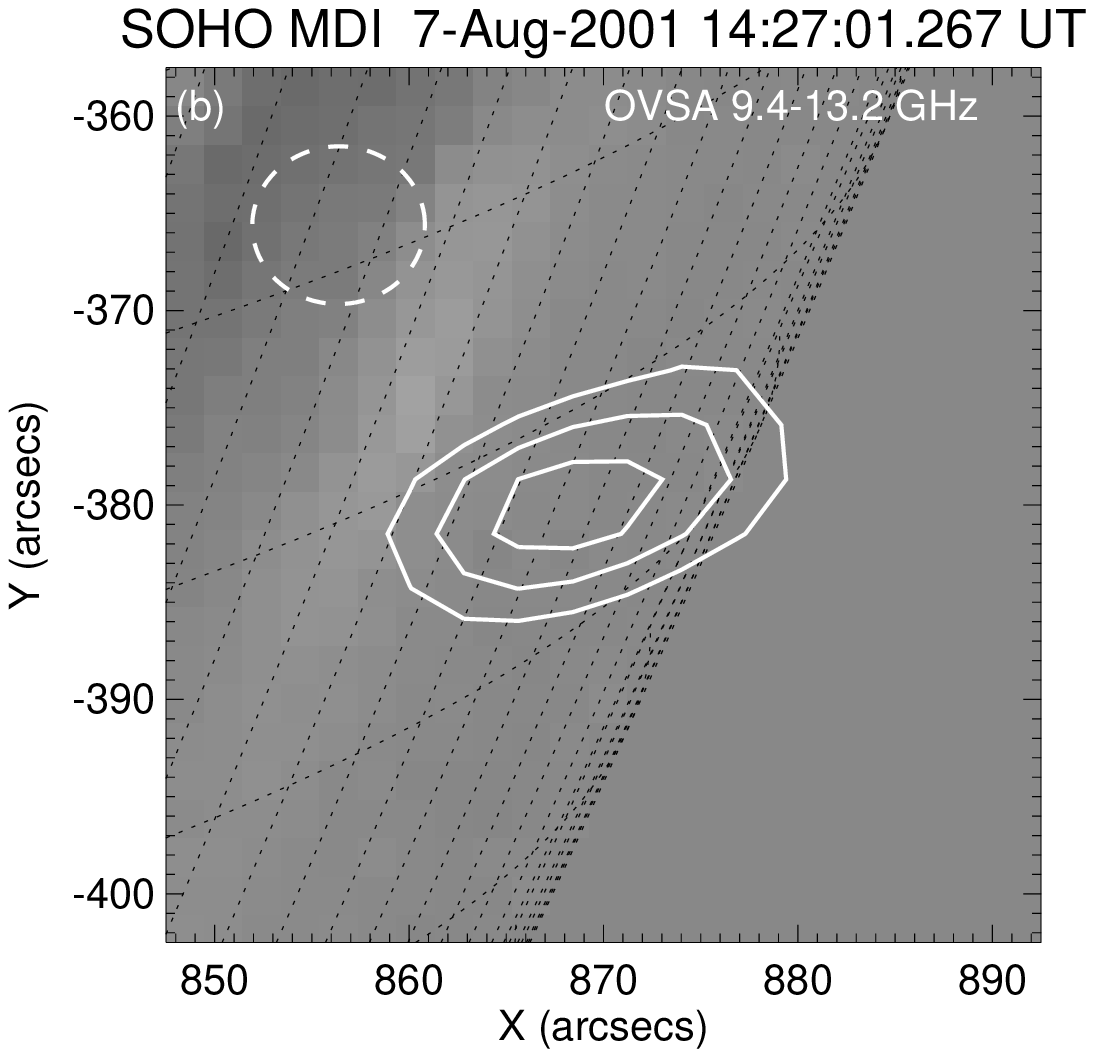}\\
\includegraphics[width=0.49\columnwidth,clip]{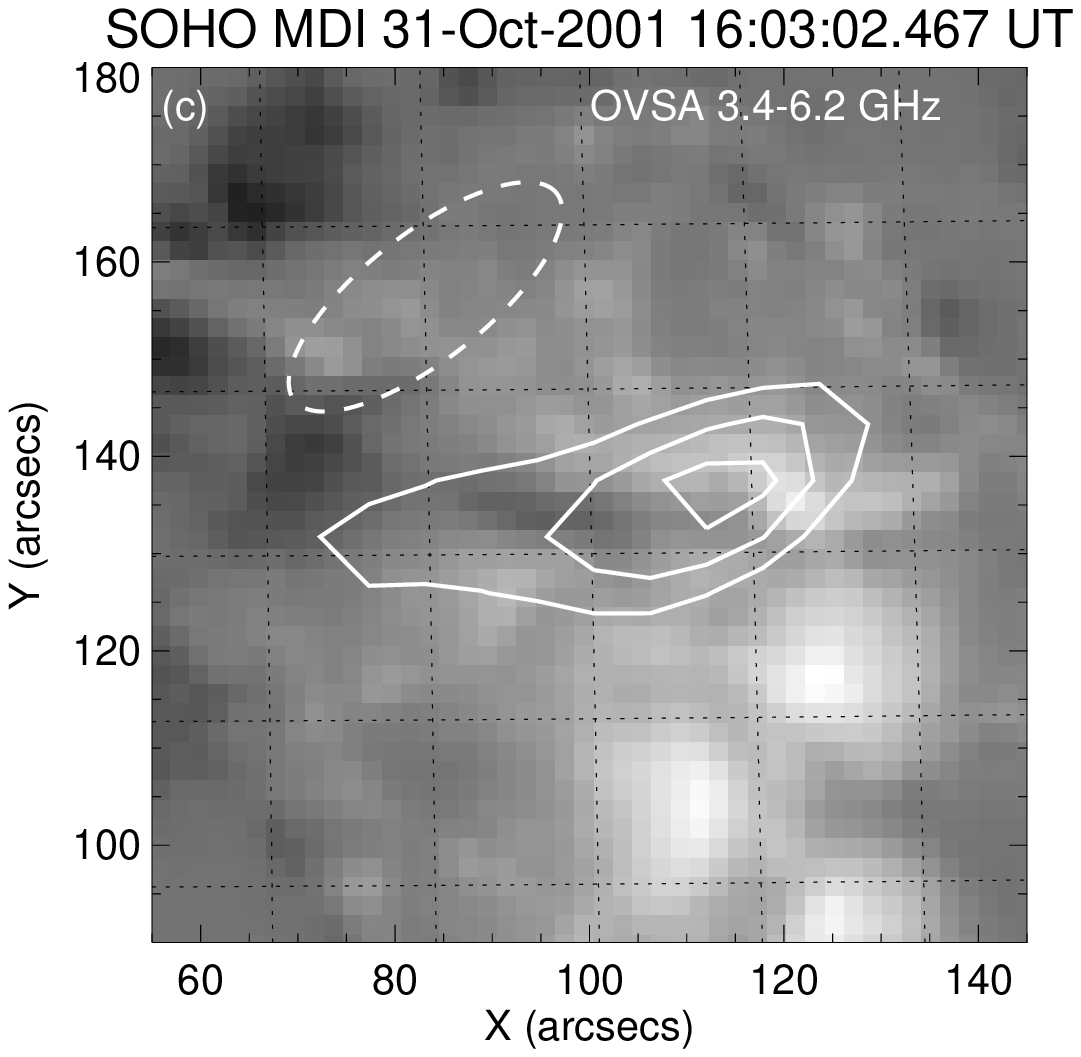}
\includegraphics[width=0.49\columnwidth,clip]{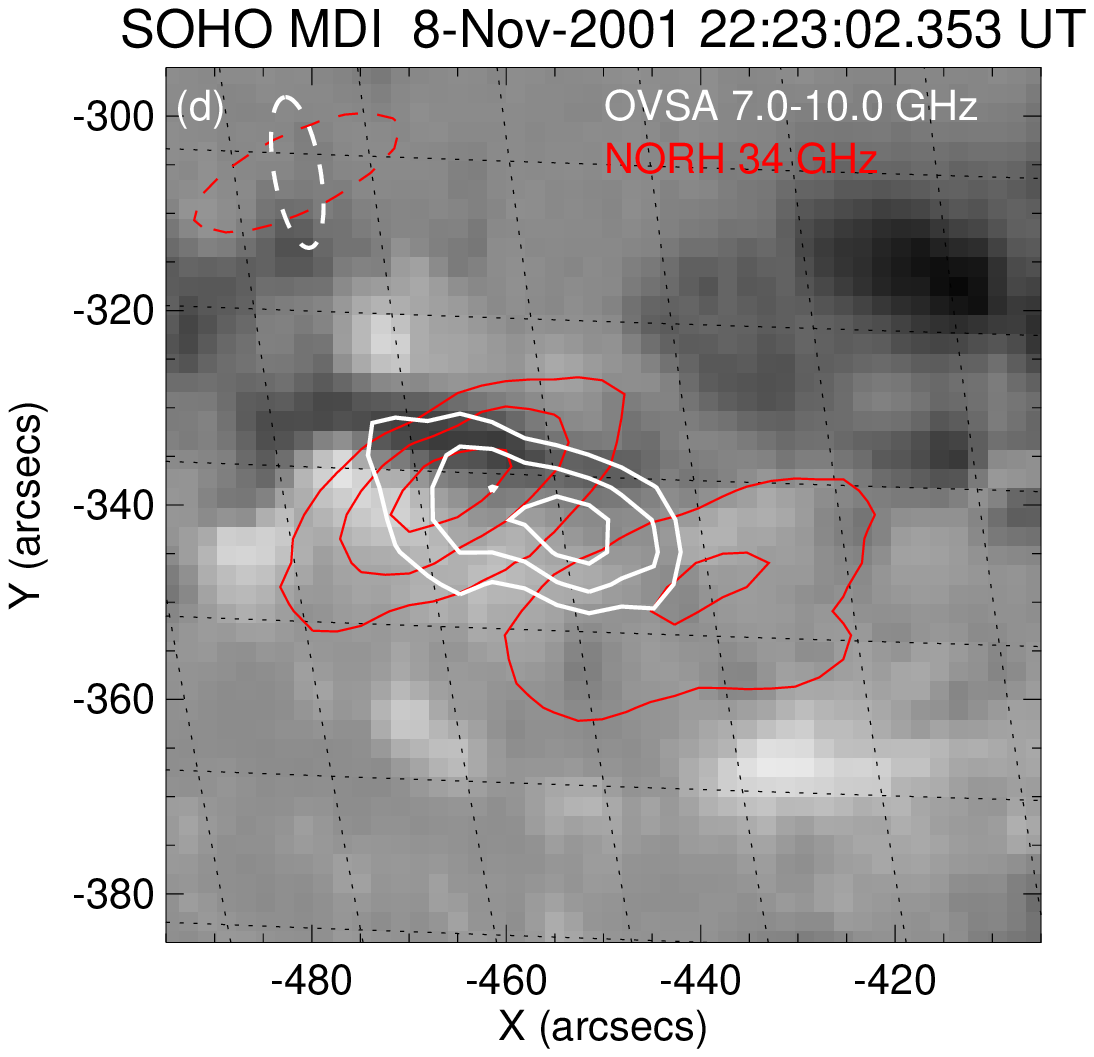}\\
\includegraphics[width=0.49\columnwidth,clip]{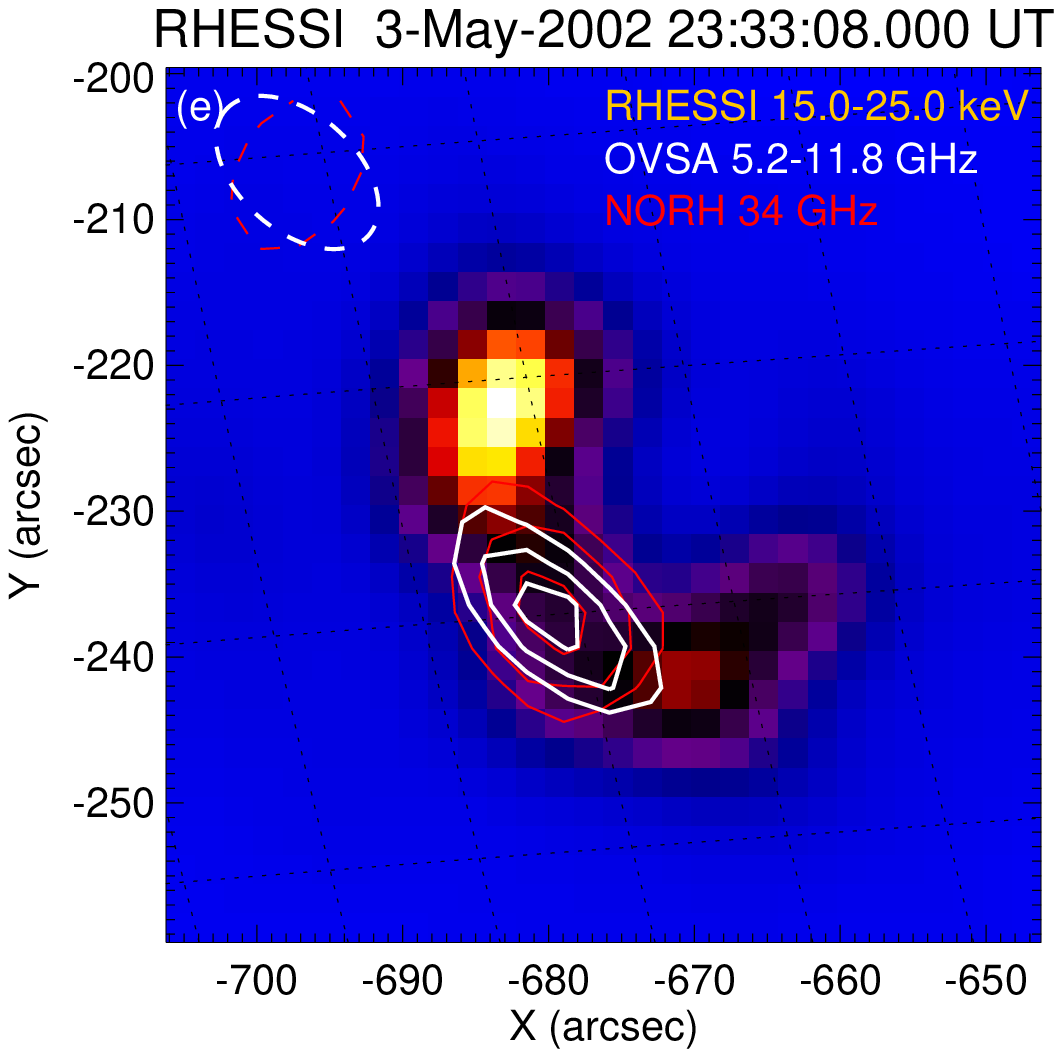}
\includegraphics[width=0.49\columnwidth,clip]{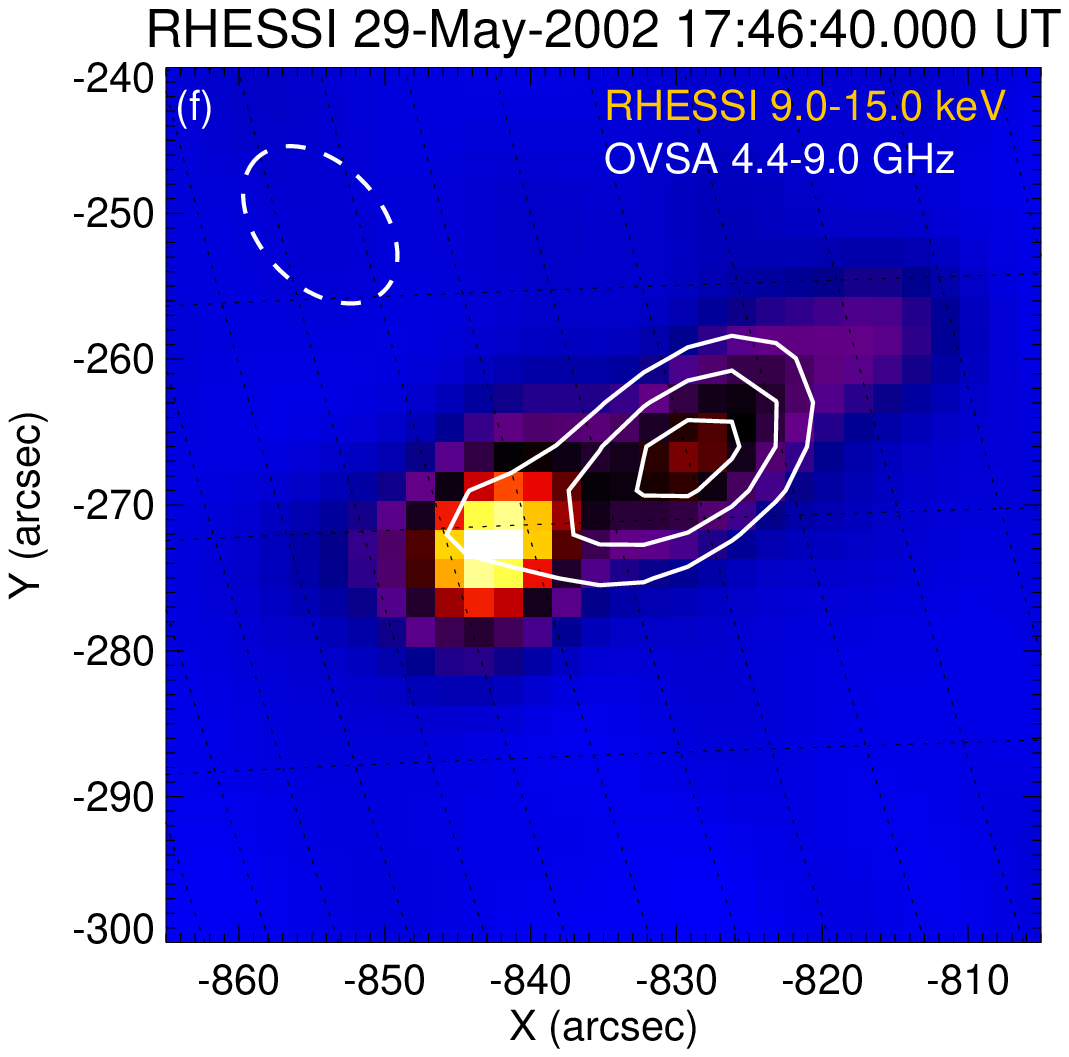}\\
\includegraphics[width=0.49\columnwidth,clip]{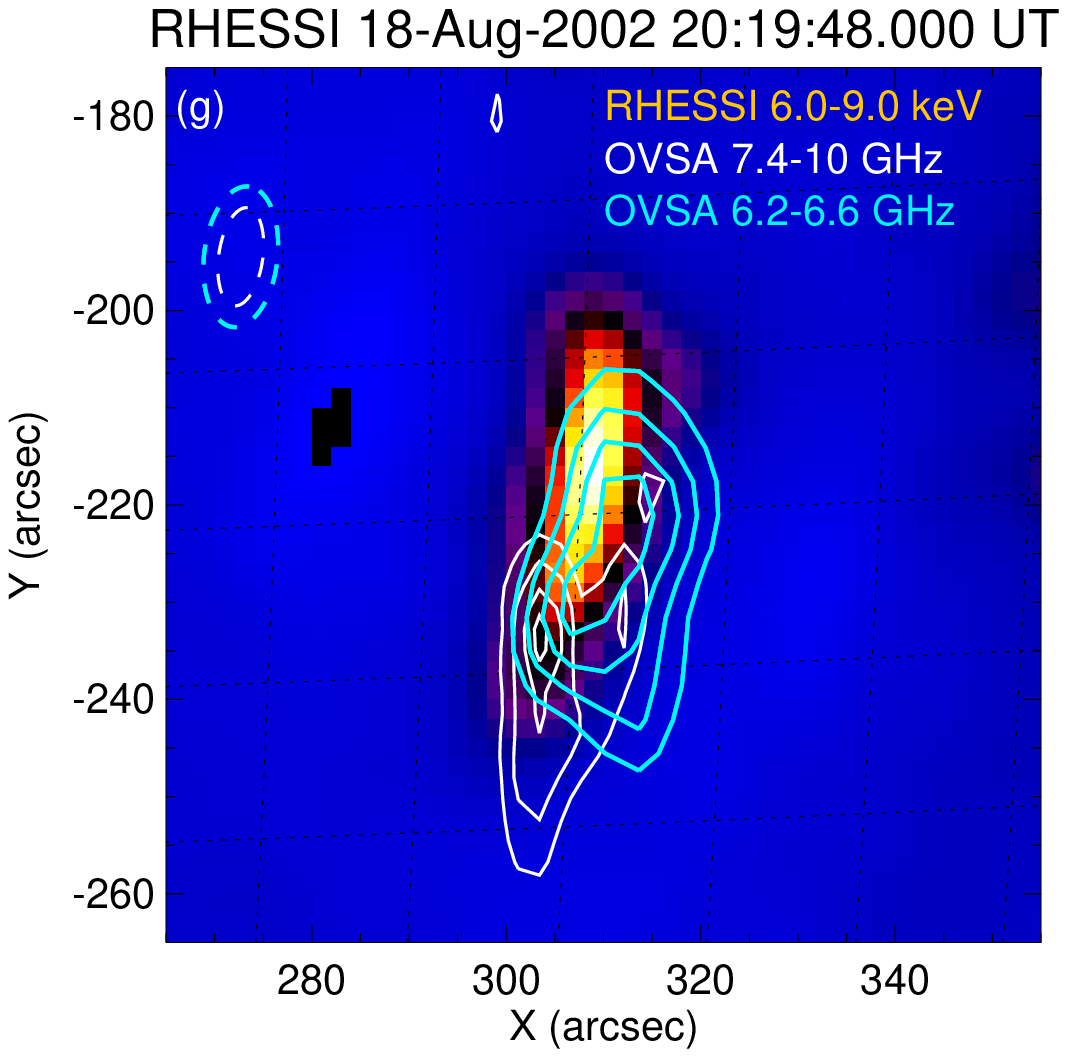}
\includegraphics[width=0.49\columnwidth,clip]{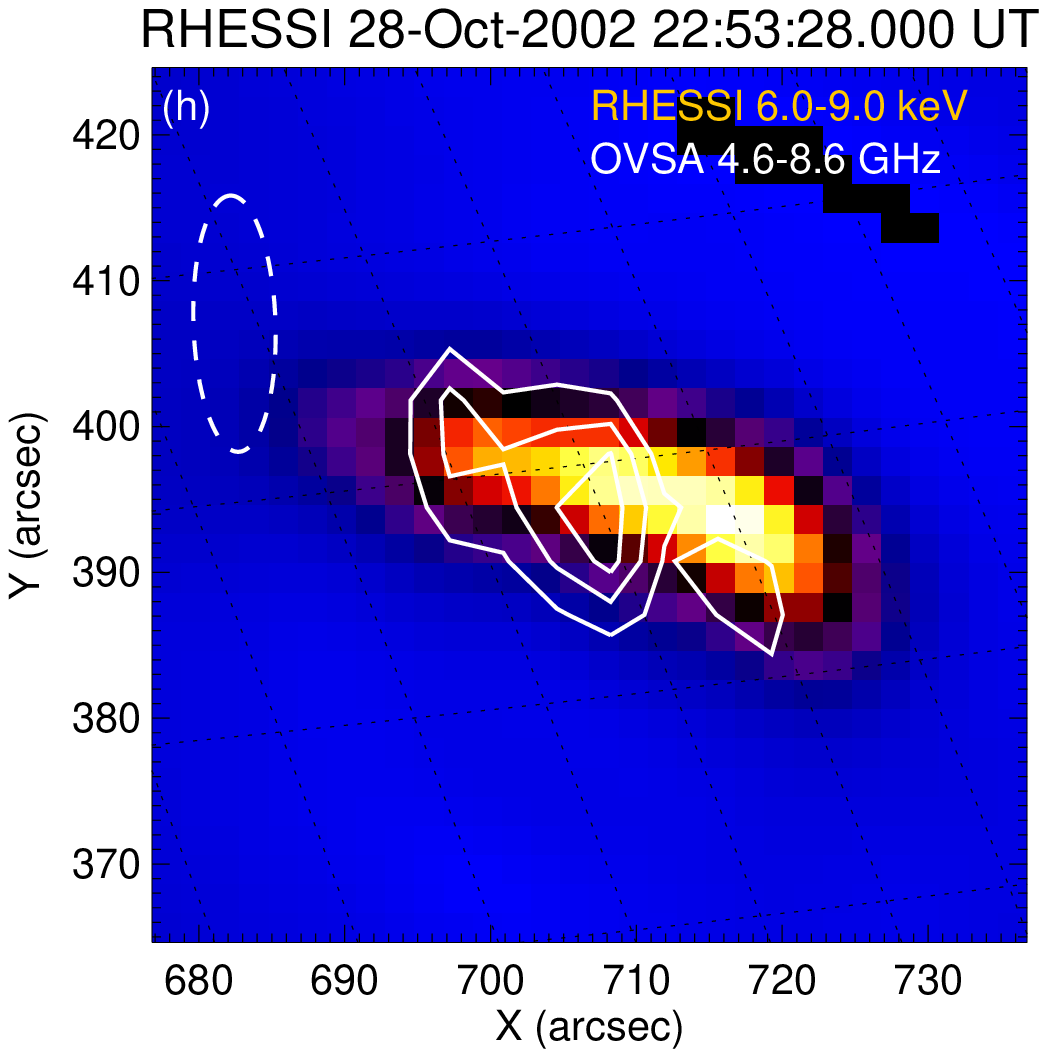}
\end{center}
\figcaption{\small \Mw\ images of the narrowband bursts on top of SoHO/MDI maps (for pre-\rhessi\ era) or on top of \rhessi\ maps. When available: NoRH images at 34~GHz are shown in red. OVSA synthesis (over spectral/temporal peak of the dynamic spectrum) map (white) obtained with the CLEAN+SELFCAL method on top of a corresponding background image. Offsets of (23'',13'') and (7'',-12'') are applied for 03-May-2002 and 29-May-2002 events, respectively. No offset is applied for 18-Aug-2002 event; main source (7.4--10.6~GHz) is shown in white, while the low-frequency (6.2--6.6~GHz) source is shown in cyan.
\label{images}}
\end{figure}

To perform OVSA imaging for the analyzed narrowband events, we created the required calibrated files for all cases where this was permitted by the data availability, quality, and completeness. Unfortunately, for many events the imaging attempt failed, due to various reasons such as data missing on one or both big antennas, the array was pointed to the wrong active region, poor quality of phase calibration observations, or the event itself could be too weak to be reliably imaged. Eventually, we selected 6 of the 21 events whose frequency-synthesis maps clearly indicated a distinct burst source (four bursts from 2001 and two from 2002---2002-May-03 and 2002-Aug-18 20:18 events).  Once the location of the burst is known, individual images are made using either the CLEAN or CLEAN+SELFCAL restoration method.
Based on the availability of joint X-ray imaging data, two additional events, 2002-May-29 and 2002-Oct-28, were included in this study, which although they permit imaging, are more ambiguous (see Fig.~\ref{images}).

The OVSA phase calibration method, based on observations of cosmic sources, normally provides an accuracy of source location limited only by the phase fluctuations of the calibrator measurement---typically amounting to a few arcsec. Nevertheless, in a few cases we find an unexplained offset of the source location relative to other data due to some as-yet unknown systematic phase shift. A striking example of a clearly wrong position is for the 2002-May-03 event. Although there is no unusual internal inconsistency in the calibration, and all imaging checks show no anomaly, the restored location of the source at all OVSA frequencies is shifted from that of the jointly observed NoRH source at 17 and 34~GHz. Accordingly, in Fig.~\ref{images} we applied an offset of ($23'',13''$), moving the OVSA source $23''$ west and $13''$ north to match the location of the NoRH source\footnote{Although the absolute source location determined by NoRH can also be incorrect (up to $10''$) because of the calibration method adopted by NoRH, especially, for strong events, in this relatively weak burst the NoRH position seems correct---nicely connecting the X-ray footpoints.}, see Fig.~\ref{images}.  Likewise, for the event of 2002-May-29 an offset of ($7'',-12''$) was required to roughly match the X-ray source morphology. A different behavior is seen in the 2002-Aug-18 20:19 event: the low-frequency (6.2--6.6~GHz) image matches the X-ray source without any offset, while the high-frequency (7.4--10.6~GHz) image is displaced southward compared to them; this frequency-dependent shift may be real, and we return to this behavior below. 

In what follows we use the imaging information to constrain the source sizes for the GS fit and also rely on the \mw-to-X-ray image relationships in interpretation of the results obtained.

\begin{figure}
\epsscale{0.98} \plotone{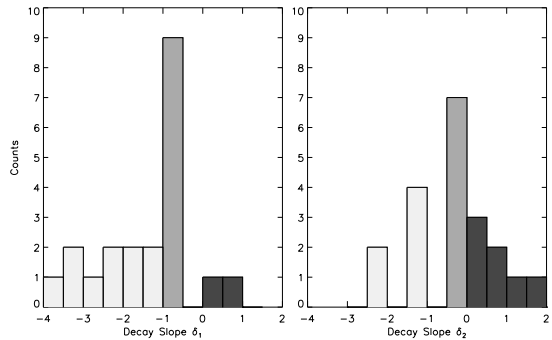}
\figcaption{\small Distributions of the  indices $\delta_{1,2}$ in the frequency dependence $\tau\propto f^{\delta_{1,2}}$, given for each individual event in Figures~\ref{decay_1_narrow} and \ref{decay_2_narrow}, for the early (left) and late (right) decay phases. Light grey--not consistent with any trapping regime; grey--strong or moderate diffusion; dark grey--weak diffusion.
 \label{Decay_slope_narrow}}
\end{figure}

\section{Radio spectral fits and comparison with X-rays}
\label{S_mw_fit}

We performed spectral fits for all events from the list, to derive \mw\ source parameters. A summary of the fit results is presented in Table~\ref{distrib_table_mw}. It is interesting that in all cases a straight-forward fit of GS plus free-free emission fits the data despite the narrow spectral shapes, i.e., no case is identified that would require an emission mechanism distinct from GS. The radio source width and depth along the line of sight are assumed to be identical and typically lie in the range $10-20''$, while the length is in the range $10-40''$. The electron spectrum is typically soft, with spectral indices $\delta\gtrsim5$, similar to what was found from the HXR spectral fits. The magnetic field is within 200-600~G. The thermal number density is quite high in many cases, $n_e \gtrsim 10^{11}$~cm$^{-3}$, in order that Razin suppression can create the observed steep low-frequency slope.  However, during times when the low-frequency slope is not particularly steep, the fit yields lower densities. It is interesting to note that the thermal number density derived from the \mw\ fit is typically an order of magnitude larger than that from the X-ray thermal fit assuming the same emitting volume.  To check this unexpected finding, we repeated the same GS+FF fitting of the \mw\ spectra, but now with the thermal number density fixed at the values derived from the X-ray fits. In most cases, an acceptable fit requires very high values of the nonthermal electron number such as $n_r\sim 10^{11}{\rm cm^{-3}} \gg n_{th}$, to create the required Razin suppression directly from the nonthermal electron density. In addition, such fits seem to require low magnetic field strength together with a very hard electron spectrum, but with an extremely low high-energy cut-off. While not formally excluded, we conclude that such special circumstances are unlikely, and instead we prefer the high thermal densities derived from the unconstrained fits, together with other fit parameters given in Table~\ref{distrib_table_mw}.

We now describe the general findings for the \mw\ fits, with exceptions indicated by the footnotes to Table~\ref{distrib_table_mw}.  For most events, we find that $E_{\max}$, the electron high-energy cutoff, is poorly constrained since the fits are not very sensitive to $E_{\max}\gtrsim 1$~MeV. When it is constrained (two cases, see $\sharp$ footnote in Table~\ref{distrib_table_mw}), it is rather small, only a few hundred keV. Although the energy spectrum of fast electrons is well determined by the fits, the derived total number of fast electrons is poorly determined, since it is biased by the arbitrarily chosen $E_{\min}=20$~keV. 
For these narrowband bursts, which require Razin suppression, the ambient plasma temperature could be 
determined from the 
free-free opacity in the low-frequency 
part of the \mw\ spectrum \citep{2007ApJ...666.1256B}. However, it is in fact poorly constrained; the fit typically drives it toward the higher end (40 MK) to avoid too strong free-free absorption at low frequencies and excessive free-free emission at high frequencies. These cases are shown by ellipsis (\dots) in the temperature column.  There are a few cases, nevertheless, where the temperature determined from the fit is low---less than or about 10~MK. In two cases ($\Uparrow$ footnote) the fit reveals an increase of the temperature with time similar to the cold flare event reported by \citet{2007ApJ...666.1256B}. In the only event confidently showing delay of the lower-frequency light curves relative to the higher-frequency light curves, 2003-Aug-06  \citep[reminiscent of the plasma heating in the cold dense flare event as in][]{2007ApJ...666.1256B}, the temperature is constrained from the fit, but no temperature increase with time is evident.  In contrast, the 2003-10-28 19:05 event does show systematic increase of the thermal number density and temperature, but does not show a significant frequency-dependent delay.

\section{Delay and decay times }
\label{S_timing_obs}

Acceleration of electrons and their transport are the two main physical processes that control the light curves of the radio and X-ray emission, including their decay and relative delays. Figure~\ref{delays_narrow} shows delays of the radio light curves relative to the light curve at the frequency showing the absolute peak flux for a given event, which is indicated by the vertical dashed line in each of the panels. We have already discussed this Figure in the context of plasma heating diagnostics and found that only very few events display the anticipated delay of the low-frequency light curves, which would have indicated plasma heating. Equally, very few events show significant delay, $>2$~s, (either positive or negative) of the high-frequency light curves: most of the events show no delay.

Study of the decay phase is somewhat more complicated because in most cases the decay phase does not follow a single exponential profile. A good model found applicable to all events is that the light curve after the peak time consists of either one or two exponential segments---one of them follows just after the peak and the other one (if present) is a slower late-decay-phase segment. The decay constants related to each segment were determined based on inspection of the logarithmic derivatives; they are plotted in Figures~\ref{decay_1_narrow} and \ref{decay_2_narrow}, respectively. In most cases the decay constants behave differently at the early and late decay phases, while in one case (08-Nov-2001) they are almost the same, indicating that there is only one time constant in this event.  In one other case (21-Oct-2003) the entire decay phase is too short to be divided onto two stages.

The decay constants $\tau(f)$ do show a prominent frequency dependence in almost all cases, which deserves closer attention. For the sake of further analysis we fit either the entire range or the high-frequency part of the decay plots by power-law functions $\tau(f)\propto f^{\delta_{1,2}}$  (thick lines); the corresponding indices $\delta_{1,2}$ for the early and late decay phases respectively are printed in each panel. Furthermore, histograms of these indices are given in Figure~\ref{Decay_slope_narrow}.


\section{Fast electron acceleration and trapping}
\label{S_acc_trap}


To facilitate further discussion let us assume that there can be two distinct coronal regions containing fast electrons---an acceleration region (A) and a trapping region (T). Further, let us assume that the volume averaged electron distribution functions in these two regions can be described by two equations:
\begin{equation}
\label{Eq_transport_Acc}
 \frac{\partial F_A(E,t)}{\partial t} + \hat{L} F_A(E, t) + \frac{F_A(E,t)}{\tau_e(E)} = 0,
\end{equation}
\begin{equation}
\label{Eq_transport_trap}
 \frac{\partial F_T(E,t)}{\partial t}  + \frac{F_T(E,t)}{\tau_T(E)} = Q(E,t).
\end{equation}
The operator $\hat{L}$ in Eq.~(\ref{Eq_transport_Acc}) for the acceleration region describes all relevant physical processes (e.g., acceleration, isotropization, mirroring etc) except particle escape from the acceleration region, which is explicitly given by the last term in Eqn~(\ref{Eq_transport_Acc}). The particle escape is explicitly treated in the relaxation time approximation.  This form is convenient because the particle loss term from the acceleration region forms the injection term for the trapped particles. In particular, if escape to interplanetary space along open field lines and precipitation down to chromospheric footpoints are both negligible, then $Q(E,t)\equiv F_A(E,t)/\tau_e(E)$, while in a general case $Q(E,t)= \eta F_A(E,t)/\tau_e(E)$, where $\eta(E)$ is an energy-dependent fraction of the particles escaping the acceleration region that feeds the trapping region; so in either case the characteristic time scale of the function $Q(E,t)$ is the same as the characteristic time scale $\tau_A(E)$ of the acceleration component $F_A(E,t)$.

Solution of Eqn~(\ref{Eq_transport_trap}) can be written down in the general form of a convolution of the injection function $Q(E,t)$ and `accumulation' described by an exponential term with an energy-dependent escape time $\tau_T(E)$ from the trapping region. In this case the angle-averaged distribution function $F_T(E,t)$ of fast electrons at the source is
\begin{equation}
\label{Eq_trapping}
 F_T(E,t)=\int_{-\infty}^t \exp\left( -\frac{t-t'}{\tau_T(E)}\right) Q(E,t')dt'.
\end{equation}
Within this concept all physical processes responsible for particle acceleration and evolution occur inside the acceleration region and are included in Eqn~(\ref{Eq_transport_Acc}), while the dominant property of the trapping region is to accumulate particles that come from the acceleration region. Importantly, the value and functional form of the electron escape time from the trapping source $\tau_T(E)$  depends on the transport mode. Solution~(\ref{Eq_trapping}) is especially convenient when the acceleration time scale $\tau_A(E)$ of the injection $Q(E,t)$ duration differs strongly from the trapping time $\tau_T(E)$.  We now derive the energy-dependent delay $\Delta t(E)$ for the two limiting cases, $\tau_A(E)\gg\tau_T(E)$ and $\tau_A(E)\ll\tau_T(E)$.

If $\tau_A(E)\gg\tau_T(E)$ then we can take $Q(E,t')$ out of the integral, which yields

\begin{equation}
\label{Eq_weak_trapping}
 F_T(E,t)\approx \tau_T(E)Q(E,t)\approx \eta(E)\tau_T(E)F_A(E,t)/\tau_e(E), $$$$ \qquad {\rm for}\ \ t\gg\tau_T.
\end{equation}
This implies that no significant trapping takes place and the time evolution of the distribution function $F_T(E,t)$ simply follows the time evolution of the acceleration function $F_A(E,t)$, although their dependencies on the energy are different because of $\eta(E)\tau_T(E)/\tau_e(E)$ factor. This also means a very small delay between the injected and trapped components at a given energy. To estimate this time delay,
we take the time derivative of Eqn~(\ref{Eq_trapping}), which yields

\begin{equation}
\label{Eq_trapping_deriv}
 F_T'(E,t)=Q(E,t) - \frac{1}{\tau_T(E)}\int_{-\infty}^t \exp\left( -\frac{t-t'}{\tau_T(E)}\right) Q(E,t')dt'
\end{equation}
and find the peak time of the trapped component at the given energy $E$ from the condition $F_T'(E,t)=0$.

In this limit of small trapping time, $\tau_A(E)\gg\tau_T(E)$, we can replace the variables $t'=t+\tau$, make an expansion $Q(E,t')\approx Q(E,t) +Q'(E,t)\tau+ Q''(E,t)\tau^2/2$, and substitute it into Eq~(\ref{Eq_trapping_deriv}), so that
\begin{equation}
\label{Eq_trapping_deriv_weakTr}
 F_T'(E,t)=-\tau_T(E)\frac{\partial Q(E,t)}{\partial t}+\tau_T^2(E)\frac{\partial^2 Q(E,t)}{\partial t^2},
\end{equation}
where the second term is needed because the first one vanishes at the injection peak, where $Q'=0$. It is convenient to adopt $Q\propto\exp(-t^2/\tau_A^2)$; then, explicit calculation of the derivatives and discarding some small terms yields an energy-dependent delay
\begin{equation}
\label{Eq_time_delay_1}
 \Delta t(E) \approx \tau_T(E),
\end{equation}
which is indeed small compared with the injection time scale $\tau_A$.

In the opposite case of efficient trapping, $\tau_A(E)\ll\tau_T(E)$, we obviously obtain
\begin{equation}
\label{Eq_strong_trapping}
 F_T(E,t)\approx\exp\left( -\frac{t}{\tau_T(E)}\right) \int_{-\infty}^{\infty}  Q(E,t')dt', $$$$ \qquad {\rm for}\ \ t\gg\tau_A .
\end{equation}
In this case the decay phase does not depend on the injection profile, but rather is solely determined by the exponential tail with the escape time $\tau_T(E)$. To estimate the delay time we have to consider Eq~(\ref{Eq_trapping_deriv}) for the case of long trapping time $\tau_A(E)\ll\tau_T(E)$. Here the exponential factor can safely be set to one, so the remaining integral is estimated as $\int_{-\infty}^{t}  Q(E,t')dt'\approx  Q(E,0)\tau_A$ if $t>\tau_A$, so
\begin{equation}
\label{Eq_trapping_deriv_strongTr}
 F_T'(E,t)\approx Q(E,t)-Q(E,0)\frac{\tau_A}{\tau_T}.
\end{equation}
Assuming an exponential decay phase of the injection profile, $Q(E,t)=Q(E,0) \exp(-t/\tau_A)$, the solution of equation $F_T'(E,t)=0$ yields\footnote{In the case of the Gaussian profile, $Q\propto\exp(-t^2/\tau_A^2)$, one has $ \Delta t(E) \approx \tau_A \ln^{1/2} \frac{\tau_T}{\tau_A}$.}

\begin{equation}
\label{Eq_time_delay_2}
 \Delta t(E) \approx \tau_A \ln \frac{\tau_T}{\tau_A}
  ,
\end{equation}
thus, the energy-dependent delay here is primarily determined by the injection duration, depending only logarithmically on $\tau_T(E)$.

\section{Microwave timing vs acceleration and trapping modes}

For our study it is important that microwave emission at a given frequency $f$ is weighted with the fast electron distribution function at a certain energy such as $f\propto E^{a}$. For example, for relativistic electrons $a=2$ so $f\propto E^2$, while for non- and moderately relativistic electrons $a\approx1-1.5$.
Therefore, studying the delay and decay times of the microwave emission as a function of frequency can be converted to diagnostics of the fast electron evolution at various energies.

Let us consider the time delays and decay constants obtained in Section~\ref{S_timing_obs} within the framework of electron acceleration and trapping presented in Section~\ref{S_acc_trap}. First, we test the widely accepted scenario that the radio light curves are formed by the electron trapping in the looptop region. Assume that the trapping time is a power-law function of the electron energy,
\begin{equation}
\label{Eq_escape_time}
 \tau_T(E) \propto E^{\alpha}
\end{equation}
and consider what ranges of $\alpha$ are available for different transport regimes---the weak, moderate, and strong diffusions; see below for the definitions. In doing so we will translate the energy dependence to the frequency dependence $\tau_T(E) \propto\tau_T(f^{1/a})$, i.e., $\tau_T(E) \propto f^{\alpha/2}$ for relativistic particles and to $\tau_T(E) \propto \tau_T(f)\propto f^{\alpha/(1-1.5)}$ for non- or moderately relativistic particles.

The weak diffusion regime implies that the particle experiences many bounces before its pitch angle changes noticeably due to angular scattering, which can be caused by either Coulomb scattering or a weak level of wave turbulence. In the former case, the escape time is specified by the Coulomb deflection time, $\tau_c \approx 0.95\cdot10^8 E_{keV}^{3/2} n_{th}^{-1}(20/\ln\Lambda)$, for nonrelativistic electrons; accordingly, the decay constant increases with frequency as $\propto f^{3/2a}$, where $a$ has been defined above. For relativistic electrons the angular scattering by the Coulomb collisions is inessential because the Coulomb energy losses is a faster process, whose time scale increases linearly with energy, $\tau_{cE}\propto E$, which translates to the microwave decay time $\propto f^{1/2}$.

The weak diffusion regime can take place even in the presence of turbulent waves if their energy density is somewhat low. In this case the escape time is specified by the particle isotropization time due to the wave-particle interaction, which differs depending on the turbulence spectrum, $\tau_{iso}\propto\beta^{1-\nu} \gamma^{2-\nu}$, where $\beta=v/c$ is the dimensionless particle velocity, $\gamma=1/\sqrt{1-\beta^2}$ is the Lorenz-factor, $\nu$ is the spectral index of the turbulence spectrum that is assumed to have a power-law form, $W_k\propto k^{-\nu}$ (such as $\int_{k_0}^{\infty}W_k dk=W_{\rm tot}$ is the total turbulence energy density). In what follows we will assume a 'classical' range of the turbulence spectral indices, $1< \nu<2$. The corresponding range of $\delta$ ($=\delta_{1,2}$, see Section~\ref{S_timing_obs}) differs for nonrelativistic and relativistic particles (see Table~\ref{T_diff_regimes}) and overall belongs to the range $-1/2 < \delta < 1/2$. Therefore, any version of the weak diffusion mode, either with Coulomb or turbulent wave scattering, implies $\delta> -1/2$.

If the turbulence level is strong enough to provide a noticeable isotropization over one bounce period the diffusion regime changes to the moderate diffusion mode. In this case the probability $P_{esc}$ to escape the loop during one bounce period is determined only by the mirror ratio $m$ of the loop, so $P_{esc}=1/m \ll 1$ for a large $m$, while the bounce period is $L/v$, i.e., $\tau=mL/v\propto v^{-1}$. Accordingly, we have $-1/2 < \delta < -1/3$ in the nonrelativistic case and $\delta=0$ in the relativistic case. Interestingly, the same ranges of $\delta$ appear for the free streaming transport of the accelerated electrons along the field lines, but the absolute value of the escape time is now smaller, $\tau\approx L/v\propto v^{-1}$.

Finally, even stronger turbulence will result in spatial diffusion (multiple changes of the propagation direction) of the electrons at the source, thus, the escape time is specified by the particle mean free path $\Lambda$, the source size $L$, and the particle velocity $v$: $\tau\sim L^2/\Lambda v$. For the same shape of the turbulence spectrum as above, this yields  $\tau\propto\beta^{\nu-3} \gamma^{\nu-2}$ and the frequency dependence of the decay constant becomes $\propto f^{(\nu-3)/2a}$ for the nonrelativistic case and $\propto f^{(\nu-2)/2}$ for the relativistic case; summary of these regimes is given in Table~\ref{T_diff_regimes}.

Let us consider what transport regime is favored by our timing analysis. Figure~\ref{Decay_slope_narrow}, left, shows the histogram of the indices $\delta_1$ of the early decay phase. Remarkably, almost half of all events (10 of 21) have $\delta_1<-1$ and so they are inconsistent with any trapping regime assuming $\tau_T\gg\tau_A$, thus their timing implies either emission from the acceleration region directly or from a trapping region with 'inefficient' trapping, $\tau_A>\tau_T$.

\begin{figure*}
\begin{center}
\includegraphics[width=0.485\textwidth,clip]{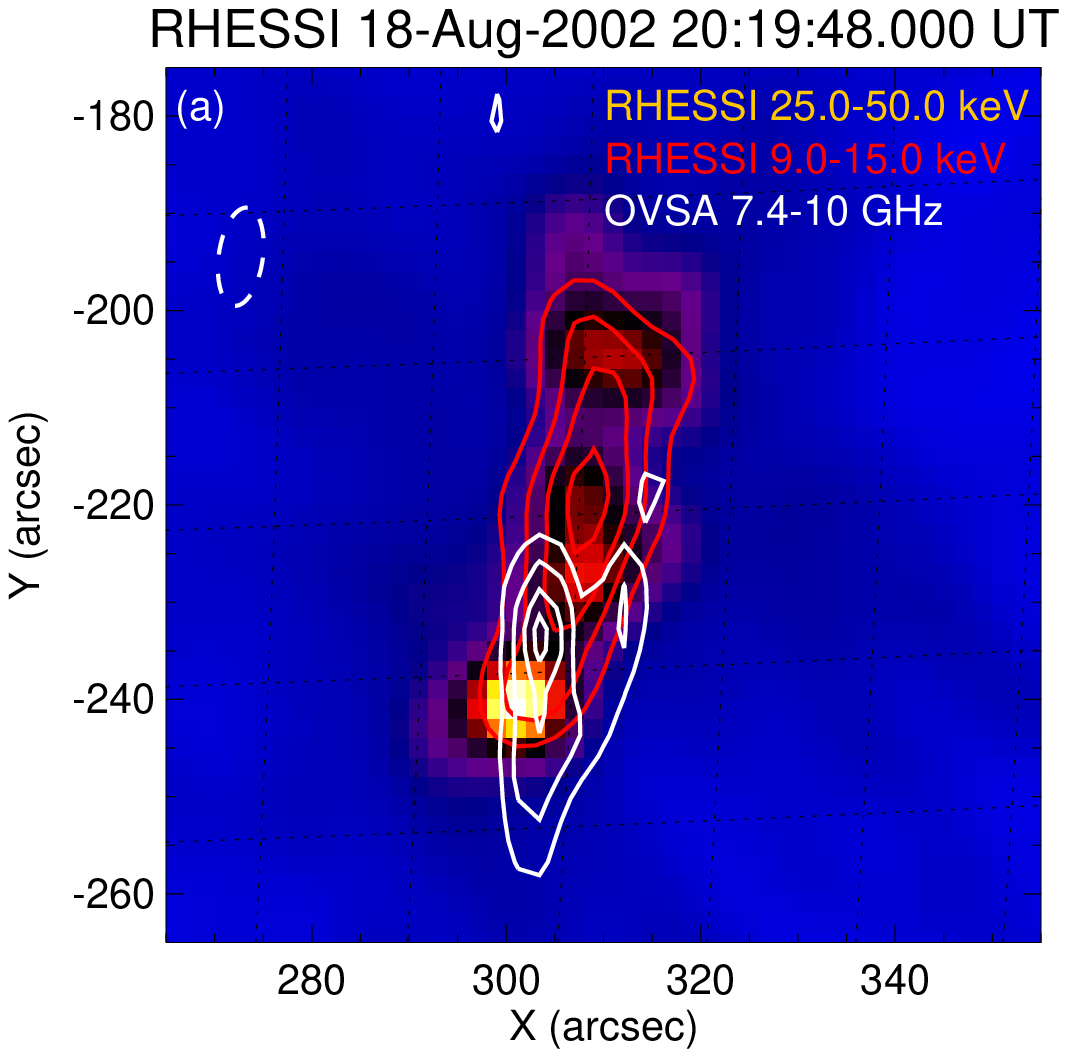}
\includegraphics[width=0.481\textwidth,clip]{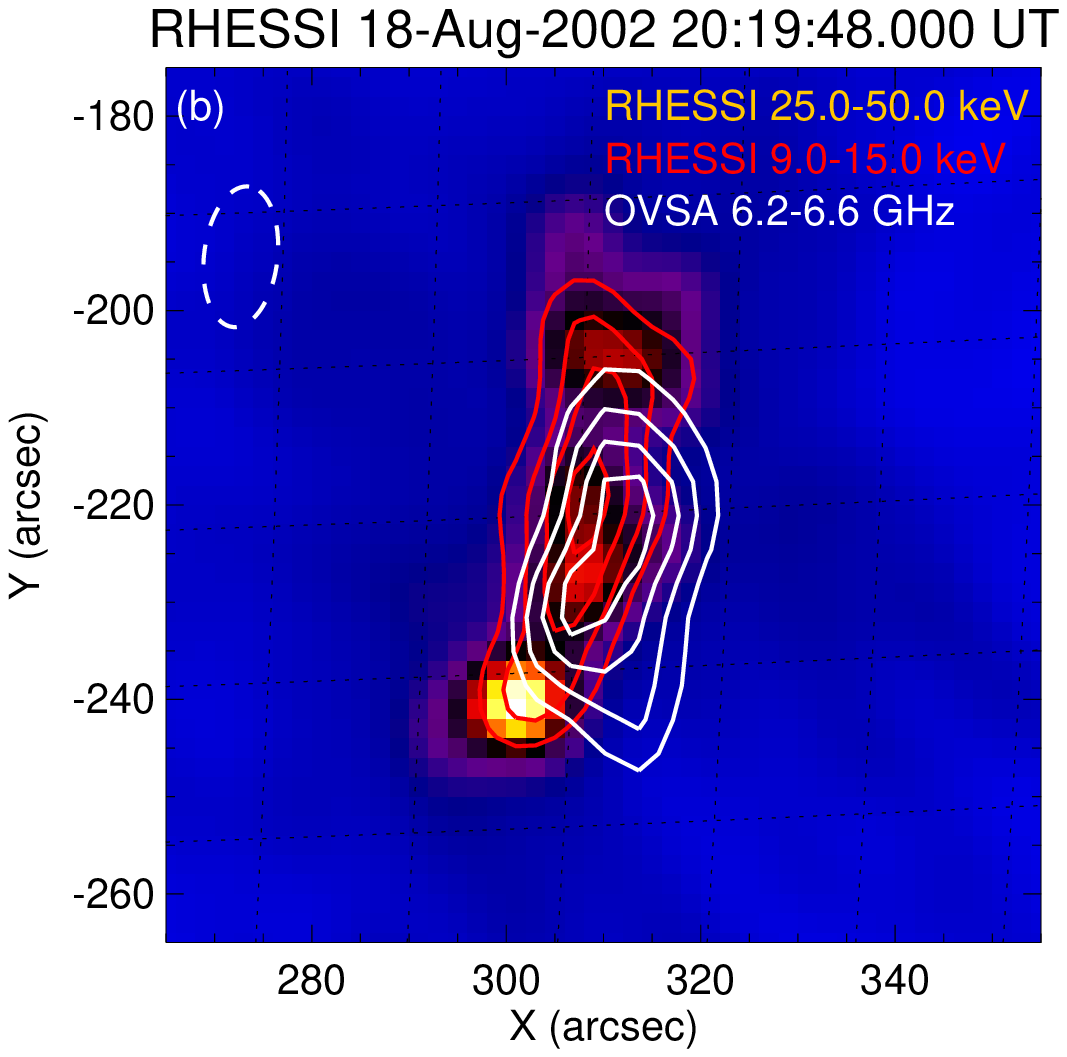}\\
\end{center}
\figcaption{\small Images for the event of 2002-Aug-18 20:19~UT. Image (25--50~keV), red contours (9--15~keV), and OVSA image at 7.4--10.6~GHz (white contours, left panel) or OVSA image at 6.2--6.6~GHz (white contours, right panel).
\label{images_20020818}}
\end{figure*}


Let us consider first the latter mode, i.e., the trapping with a short trapping time.
According to Eqn~(\ref{Eq_weak_trapping}), in this case the distribution function of the radio emitting electrons is a product of the injection function and the (small) trapping time. In this case the energy dependence of the trapping time may be deduced from the time delay between the microwave light curves at different frequencies according to Eqn~(\ref{Eq_time_delay_1}). However, no event from this group shows any noticeable time delay, see Figure~\ref{delays_narrow}, thus, only very special trapping regimes with energy-independent trapping time can be consistent with the data. Furthermore, the second equality in Eqn~(\ref{Eq_weak_trapping}) tells that the electron distribution functions in the trapping and acceleration regions are proportional to each other with the factor $ \eta(E)\tau_T(E)/\tau_e(E)$, which is reasonably small provided that $\tau_T(E)$ is small. Stated another way, the number of fast electrons in the acceleration region is larger or comparable with that at the trapping regions; thus, if we observe the trapping region, the acceleration region must also be observed as a distinct source (unless  the magnetic field and plasma density are both small at the acceleration region), which is not the case. We have to conclude that having a trapping source dominating the microwave emission  is unlikely for these 10 of 21 cases.


The other (11 of 21) events do show consistency with one of the trapping regime, although most of them (9 of 11) are only consistent with the strong diffusion mode or, marginally, with the moderate diffusion mode\footnote{The slope is also consistent with the free-steaming mode; however, the absolute values of the decay time constant are larger than 1~s and can hardly be consistent with the free-streaming mode.}, which both  imply a highly enhanced turbulence level at the source, and GS emission produced by non- or moderately relativistic electrons.
However, given the high densities of the sources, $\gtrsim10^{11}$~cm$^{-3}$ (see Table), the collisional energy losses of the non-relativistic electrons must be be essential on the burst time scale, which would manifest itself, in particular, in the spectrum flattening that is not observed. We have to conclude that the combination of the data and data products does not favor any trapping mode in the majority of these narrowband events and, thus, implies that we observe here microwave emission from directly the acceleration region. In contrast, the timing of the two remaining events is consistent with a trapping mode with the trapping time weakly rising with frequency (i.e. on energy), i.e., the weak diffusion on the turbulent waves.


Note that the distribution of the late decay time constants is markedly different from that of the early decay. Formally, many of them are consistent with one or another trapping mode. The situation here is less clean, however, because the flux levels are lower, the light curves are more complex (in some cases they are just the continuation of the early decay), and no independent delay information can be obtained for this late decay phase. Overall, we can conclude that a trapping mode does not contradict the late decay timing for most of the events.

%
%
%
%
%

\section{Discussion}

In this study we identified a subclass of solar flares whose distinctive feature is quite a  narrowband continuum spectrum observed in
microwave range with OVSA. Similar radio bursts would be missing from single
frequency instruments with large  gaps between spectral channels. We
found that such narrow spectra are formed in dense coronal sources under the action of the
Razin-effect and the free-free absorption, while the gyrosynchrotron
self-absorption played a minor role in most of these events.

Analysis of the timing and spectral evolution of these events suggests that no significant trapping of the fast electrons happens at the main phases of these bursts, rather  microwave emission originates in the acceleration region directly. These acceleration regions are characterized by relatively strong magnetic fields and high densities derived from the GS fits to the \mw\ data. The high-density coronal loops are known to produce significant X-ray emission. From this perspective, it is striking that the thermal number density derived from the thermal part of the HXR  fits is an order of magnitude or more smaller than that derived from the \mw\ fits. This implies that (i) the \mw\ and HXR  emission come from distinct volumes with different number densities and different magnetic fields, (ii) the denser thermal plasma in the \mw\ source is colder than that in the thermal HXR source; otherwise, the denser source would dominate the thermal HXR  emission, and (iii) the magnetic field is smaller at the soft X-ray source than in the \mw\ source; otherwise, the volume producing the SXR emission would dominate the radio emission. We propose that the radio sources in our sample of the narrowband \mw\  bursts are at lower heights than the accompanying SXR sources; otherwise, it is difficult to obtain the properties (ii) and (iii) itemized above. 

Perhaps, the simplest model containing two distinct volumes is a flare involving two (interacting) magnetic flux tubes: one of them is the dense loop with a strong magnetic field seen in the \mw\ emission and identified as the flare acceleration region, while the other one is a more tenuous flux tube with a smaller magnetic field seen in the thermal HXR emission. The presence of more than one loop in many of our events is confirmed by the already mentioned complexity of the HXR  source morphology, which often shows a few distinct sources (footpoints?), especially at the highest energy channel (25--50~keV). In such cases the dominant \mw\ source should spatially correlate with the high-energy HXR emission. A direct test of this model could be performed based on detailed HXR  and \mw\ image comparison at various energies/frequencies, which could reveal a multi-loop morphology; however, we only have joint \rhessi\ and OVSA imaging data for a few events from the list, so the comprehensive analysis could not be performed.

Fortunately, for the relatively strong \mw\ event of 18-Aug-2002~20:19~UT, conclusive information can be obtained. Figure~\ref{images_20020818} shows \rhessi\ 25--50~keV image demonstrating three distinct sources, presumably---footpoints, on which thermal X-ray emission at 9-15~keV is shown in red contours. This low-energy, presumably coronal, source connects two most distant HXR footpints. On the contrary, the high-frequency (7.4--10.6~GHz) image (white contours) synthesized for the \mw\ spectral peak range is located between two close southern footpoints, highlighting a smaller coronal loop that produces most of the \mw\ emission from the flare. The right panel of this plot shows a lower-frequency (6.2--6.6~GHz) source highlighting a bigger loop that connects the most distant footpoints alike the thermal X-ray source. The only viable reason for the displacement between the high- and low- frequency sources is that the spectrum of the main source dies away quickly at the lower frequency, so the emission from an overall weaker source becomes dominant at reasonably low frequencies. The lower-frequency GS source, which nearly coincides with the thermal X-ray source, implies a weaker magnetic field in that source in full agreement with our interpretation of the data. Finally, the lack of the \mw\ emission from the second magnetic flux tube implies that the trapping is inefficient there.




\section{Conclusions}

In this study we identified a new subclass of solar flares, which is characterized by a narrowband spectrum of \mw\ emission that comes likely from directly the acceleration region. The analysis performed implies that energy release in these flares occurred due to interaction between at least two distinct loops. One of the loops is a compact dense loop with a reasonably strong magnetic field, which is observed as a narrowband \mw\ source, whose physical parameters are determined through the spectral fit. The other loop is a more tenuous bigger loop, with a smaller magnetic field, which is seen in SXR but barely or not seen in the \mw. Study of the detailed morphology of such flares has to await a better quality imaging radio observations.

%
%
%
%

\acknowledgments  
This work was supported in part by NSF grants  AGS-1250374 and AGS-1262772 and NASA grant NNX14AC87G to New Jersey
Institute of Technology and STFC Consolidated grant project "X-ray and radio diagnostics of energetic solar flare processes," to University of Glasgow.



\bibliographystyle{apj}

\bibliography{fleishman,refs_rhessi,NarrowBandPaper} %

\begin{thebibliography}{41}
\expandafter\ifx\csname natexlab\endcsname\relax\def\natexlab#1{#1}\fi

\bibitem[{{Bastian} {et~al.}(1998{\natexlab{a}}){Bastian}, {Benz}, \&
  {Gary}}]{Bastian_etal_1998}
{Bastian}, T.~S., {Benz}, A.~O., \& {Gary}, D.~E. 1998{\natexlab{a}}, \araa,
  36, 131

\bibitem[{{Bastian} {et~al.}(1998{\natexlab{b}}){Bastian}, {Benz}, \&
  {Gary}}]{1998ARA&A..36..131B}
---. 1998{\natexlab{b}}, \araa, 36, 131

\bibitem[{{Bastian} {et~al.}(2007){Bastian}, {Fleishman}, \&
  {Gary}}]{2007ApJ...666.1256B}
{Bastian}, T.~S., {Fleishman}, G.~D., \& {Gary}, D.~E. 2007, \apj, 666, 1256

\bibitem[{{Dulk} \& {Marsh}(1982)}]{Dulk_Marsh_1982}
{Dulk}, G.~A. \& {Marsh}, K.~A. 1982, \apj, 259, 350

\bibitem[{{Fleishman} \& {Kontar}(2010)}]{2010ApJ...709L.127F}
{Fleishman}, G.~D. \& {Kontar}, E.~P. 2010, \apjl, 709, L127

\bibitem[{{Fleishman} {et~al.}(2011){Fleishman}, {Kontar}, {Nita}, \&
  {Gary}}]{Fl_etal_2011}
{Fleishman}, G.~D., {Kontar}, E.~P., {Nita}, G.~M., \& {Gary}, D.~E. 2011,
  \apjl, 731, L19

\bibitem[{{Fleishman} {et~al.}(2013){Fleishman}, {Kontar}, {Nita}, \&
  {Gary}}]{Fl_etal_2013}
---. 2013, \apj, 768, 190

\bibitem[{{Fleishman} {et~al.}(2015){Fleishman}, {Nita}, \&
  {Gary}}]{Fl_etal_2015}
{Fleishman}, G.~D., {Nita}, G.~M., \& {Gary}, D.~E. 2015, \apj, 802, 122

\bibitem[{{Gary} \& {Hurford}(1989)}]{Gary_Hurford_1989}
{Gary}, D.~E. \& {Hurford}, G.~J. 1989, \apj, 339, 1115

\bibitem[{{Gary} \& {Hurford}(1990)}]{1990ApJ...361..290G}
---. 1990, \apj, 361, 290

\bibitem[{{Gary} \& {Hurford}(1994)}]{1994ApJ...420..903G}
---. 1994, \apj, 420, 903

\bibitem[{{Gim{\'e}nez de Castro} {et~al.}(2006){Gim{\'e}nez de Castro},
  {Costa}, {Silva}, {Sim{\~o}es}, {Correia}, \& {Magun}}]{2006A&A...457..693G}
{Gim{\'e}nez de Castro}, C.~G., {Costa}, J.~E.~R., {Silva}, A.~V.~R.,
  {Sim{\~o}es}, P.~J.~A., {Correia}, E., \& {Magun}, A. 2006, \aap, 457, 693

\bibitem[{{Gim{\'e}nez de Castro} {et~al.}(2013){Gim{\'e}nez de Castro},
  {Cristiani}, {Sim{\~o}es}, {Mandrini}, {Correia}, \&
  {Kaufmann}}]{2013SoPh..284..541G}
{Gim{\'e}nez de Castro}, C.~G., {Cristiani}, G.~D., {Sim{\~o}es}, P.~J.~A.,
  {Mandrini}, C.~H., {Correia}, E., \& {Kaufmann}, P. 2013, \solphys, 284, 541

\bibitem[{{Guidice} \& {Castelli}(1975)}]{1975SoPh...44..155G}
{Guidice}, D.~A. \& {Castelli}, J.~P. 1975, \solphys, 44, 155

\bibitem[{{Guo} {et~al.}(2012){Guo}, {Emslie}, {Kontar}, {Benvenuto},
  {Massone}, \& {Piana}}]{2012A&A...543A..53G}
{Guo}, J., {Emslie}, A.~G., {Kontar}, E.~P., {Benvenuto}, F., {Massone}, A.~M.,
  \& {Piana}, M. 2012, \aap, 543, A53

\bibitem[{{Holman} {et~al.}(2011){Holman}, {Aschwanden}, {Aurass}, {Battaglia},
  {Grigis}, {Kontar}, {Liu}, {Saint-Hilaire}, \&
  {Zharkova}}]{2011SSRv..159..107H}
{Holman}, G.~D., {Aschwanden}, M.~J., {Aurass}, H., {Battaglia}, M., {Grigis},
  P.~C., {Kontar}, E.~P., {Liu}, W., {Saint-Hilaire}, P., \& {Zharkova}, V.~V.
  2011, \ssr, 159, 107

\bibitem[{{Hurford} {et~al.}(1984){Hurford}, {Read}, \&
  {Zirin}}]{1984SoPh...94..413H}
{Hurford}, G.~J., {Read}, R.~B., \& {Zirin}, H. 1984, \solphys, 94, 413

\bibitem[{{Jeffrey} {et~al.}(2014){Jeffrey}, {Kontar}, {Bian}, \&
  {Emslie}}]{2014ApJ...787...86J}
{Jeffrey}, N.~L.~S., {Kontar}, E.~P., {Bian}, N.~H., \& {Emslie}, A.~G. 2014,
  \apj, 787, 86

\bibitem[{{Kaufmann}(2012)}]{2012snc..book...61K}
{Kaufmann}, P. {Observations of Solar Flares from GHz to THz Frequencies}, ed.
  V.~N. {Obridko}, K.~{Georgieva}, \& Y.~A. {Nagovitsyn}, 61

\bibitem[{{Klopf} {et~al.}(2014){Klopf}, {Kaufmann}, {Raulin}, \&
  {Szpigel}}]{2014ApJ...791...31K}
{Klopf}, J.~M., {Kaufmann}, P., {Raulin}, J.-P., \& {Szpigel}, S. 2014, \apj,
  791, 31

\bibitem[{{Kontar} {et~al.}(2011{\natexlab{a}}){Kontar}, {Brown}, {Emslie},
  {Hajdas}, {Holman}, {Hurford}, {Ka{\v s}parov{\'a}}, {Mallik}, {Massone},
  {McConnell}, {Piana}, {Prato}, {Schmahl}, \&
  {Suarez-Garcia}}]{2011SSRv..159..301K}
{Kontar}, E.~P., {Brown}, J.~C., {Emslie}, A.~G., {Hajdas}, W., {Holman},
  G.~D., {Hurford}, G.~J., {Ka{\v s}parov{\'a}}, J., {Mallik}, P.~C.~V.,
  {Massone}, A.~M., {McConnell}, M.~L., {Piana}, M., {Prato}, M., {Schmahl},
  E.~J., \& {Suarez-Garcia}, E. 2011{\natexlab{a}}, \ssr, 159, 301

\bibitem[{{Kontar} {et~al.}(2011{\natexlab{b}}){Kontar}, {Hannah}, \&
  {Bian}}]{2011ApJ...730L..22K}
{Kontar}, E.~P., {Hannah}, I.~G., \& {Bian}, N.~H. 2011{\natexlab{b}}, \apjl,
  730, L22

\bibitem[{{Kontar} {et~al.}(2006){Kontar}, {MacKinnon}, {Schwartz}, \&
  {Brown}}]{2006A&A...446.1157K}
{Kontar}, E.~P., {MacKinnon}, A.~L., {Schwartz}, R.~A., \& {Brown}, J.~C. 2006,
  \aap, 446, 1157

\bibitem[{{Krucker} {et~al.}(2010){Krucker}, {Hudson}, {Glesener}, {White},
  {Masuda}, {Wuelser}, \& {Lin}}]{2010ApJ...714.1108K}
{Krucker}, S., {Hudson}, H.~S., {Glesener}, L., {White}, S.~M., {Masuda}, S.,
  {Wuelser}, J., \& {Lin}, R.~P. 2010, \apj, 714, 1108

\bibitem[{{Kundu} {et~al.}(2001){Kundu}, {White}, {Shibasaki}, {Sakurai}, \&
  {Grechnev}}]{Kundu_etal_2001}
{Kundu}, M.~R., {White}, S.~M., {Shibasaki}, K., {Sakurai}, T., \& {Grechnev},
  V.~V. 2001, \apj, 547, 1090

\bibitem[{{Lee} \& {Gary}(2000)}]{Lee_Gary_2000}
{Lee}, J. \& {Gary}, D.~E. 2000, \apj, 543, 457

\bibitem[{{Lee} {et~al.}(2000){Lee}, {Gary}, \&
  {Shibasaki}}]{2000ApJ...531.1109L}
{Lee}, J., {Gary}, D.~E., \& {Shibasaki}, K. 2000, \apj, 531, 1109

\bibitem[{{Lee} {et~al.}(1994){Lee}, {Gary}, \& {Zirin}}]{1994SoPh..152..409L}
{Lee}, J.~W., {Gary}, D.~E., \& {Zirin}, H. 1994, \solphys, 152, 409

\bibitem[{{Lin} {et~al.}(2002){Lin}, {Dennis}, {Hurford},
  {et~al.}}]{2002SoPh..210....3L}
{Lin}, R.~P., {Dennis}, B.~R., {Hurford}, G.~J., {et~al.} 2002, \solphys, 210,
  3

\bibitem[{{Melnikov}(1994)}]{Melnikov_1994}
{Melnikov}, V.~F. 1994, Radiophysics and Quantum Electronics, 37, 557

\bibitem[{{Melnikov} {et~al.}(2008){Melnikov}, {Gary}, \&
  {Nita}}]{2008SoPh..253...43M}
{Melnikov}, V.~F., {Gary}, D.~E., \& {Nita}, G.~M. 2008, \solphys, 253, 43

\bibitem[{{Melnikov} \& {Magun}(1998)}]{Meln_Magun_1998}
{Melnikov}, V.~F. \& {Magun}, A. 1998, \solphys, 178, 153

\bibitem[{{Melnikov} {et~al.}(2002){Melnikov}, {Shibasaki}, \&
  {Reznikova}}]{melnikov_etal_2002}
{Melnikov}, V.~F., {Shibasaki}, K., \& {Reznikova}, V.~E. 2002, \apjl, 580,
  L185

\bibitem[{{Nakajima} {et~al.}(1985){Nakajima}, {Sekiguchi}, {Sawa}, {Kai}, \&
  {Kawashima}}]{1985PASJ...37..163N}
{Nakajima}, H., {Sekiguchi}, H., {Sawa}, M., {Kai}, K., \& {Kawashima}, S.
  1985, \pasj, 37, 163

\bibitem[{{Nita} {et~al.}(2014){Nita}, {Fleishman}, \&
  {Gary}}]{2014AAS...22421845N}
{Nita}, G.~M., {Fleishman}, G.~D., \& {Gary}, D.~E. 2014, in American
  Astronomical Society Meeting Abstracts, Vol. 224, American Astronomical
  Society Meeting Abstracts \#224, \#218.45

\bibitem[{{Nita} {et~al.}(2004){Nita}, {Gary}, \& {Lee}}]{2004ApJ...605..528N}
{Nita}, G.~M., {Gary}, D.~E., \& {Lee}, J. 2004, \apj, 605, 528

\bibitem[{{Stahli} {et~al.}(1989){Stahli}, {Gary}, \&
  {Hurford}}]{1989SoPh..120..351S}
{Stahli}, M., {Gary}, D.~E., \& {Hurford}, G.~J. 1989, \solphys, 120, 351

\bibitem[{{Stahli} {et~al.}(1990){Stahli}, {Gary}, \&
  {Hurford}}]{1990SoPh..125..343S}
---. 1990, \solphys, 125, 343

\bibitem[{{Veronig} \& {Brown}(2004)}]{2004ApJ...603L.117V}
{Veronig}, A.~M. \& {Brown}, J.~C. 2004, \apjl, 603, L117

\bibitem[{{Xu} {et~al.}(2008){Xu}, {Emslie}, \&
  {Hurford}}]{2008ApJ...673..576X}
{Xu}, Y., {Emslie}, A.~G., \& {Hurford}, G.~J. 2008, \apj, 673, 576

\bibitem[{{Zaitsev} {et~al.}(2014){Zaitsev}, {Stepanov}, \&
  {Kaufmann}}]{2014SoPh..289.3017Z}
{Zaitsev}, V.~V., {Stepanov}, A.~V., \& {Kaufmann}, P. 2014, \solphys, 289,
  3017

\end{thebibliography}

\begin{deluxetable*}{llllllll}
\tablecolumns{11}
\tablewidth{0pc}
\tablecaption{\label{distrib_table_x}Flare parameters derived from fitting of spatially integrated X-ray spectrum.}
\tablehead{\colhead{Flare} & \colhead{$A_{6-9 keV}$} & \colhead{$A_{9-15 keV}$} &   \colhead{$\frac{EM}{10^{49}~{\rm cm}^{-3}}$} & \colhead{$T$, MK}  & \colhead{$\frac{N_{e}}{10^{35}~{\rm s}^{-1}}$} & \colhead{$E_{min}$, keV} & \colhead{$\delta_{nth,X}$}}
\startdata
03-May-2002 23:32 & 120 &  176 &    $0.016$  & $18.1$ & $7.31$ & $17.5$ & $5.1$\\
21-May-2002 18:25 & 172 & 176 &     $0.0007$  & $26$ & $0.12$ & $22.3$ & $4.7$\\
29-May-2002 17:46 & 92 & 116 &      $0.0023$  & $20.9$ & $0.7$ & $18.2$ & $5.1$\\
31-Jul-2002 23:27 & 328 & 468 &     $0.0036$  & $23.3$ & $0.42$ & $18.6$ & $6.0$\\
18-Aug-2002 20:18 & 260 & 360 &     $0.15$  & $17.1$ & $20$ & $14.8$ & $7.3$\\
18-Aug-2002 22:28 & 228 & 380 &     $0.027$  & $20.3$ & $2.5$ & $17.8$ & $7.3$\\
20-Aug-2002 22:08 & 76 & 92 &       $0.011$  & $21.6$ & $0.8$ & $17.7$ & $6.1$\\
20-Aug-2002 22:15 & 184 & 136 &     $0.0001$  & $30.2$ & $0.016$ & $21.8$ & $4.9$\\
28-Oct-2002 22:51 & 248 & 276 &     $0.016$  & $12.2$ & $0.5$ & $13.4$ & $10.3$\\
29-Oct-2002 16:26 & 396 & 388 &     $0.14$  & $20.6$ & $0.87$ & $21.1$ & $10.0$\\
06-Jul-2003 20:02 & 212 & 280 &     $0.015$  & $15.2$ & $3.41$ & $15.8$ & $6.4$\\
21-Oct-2003 20:20 & 580 & 676 &   $0.0087$  & $19.7$ & $0.058$ & $21.8$ & $4.48$\\
\enddata
\end{deluxetable*}

\begin{deluxetable*}{lllllllll}
\tablecolumns{9}
\tablewidth{0pc}
\tablecaption{\label{distrib_table_mw} Flare parameters derived from microwave fit}
\tablehead{\colhead{Flare} & \colhead{sizes} & \colhead{$B$, G} & \colhead{$\left(\frac{n_{th}}{{10^{11}~{\rm cm}^{-3}}}\right)$\tablenotemark{$\bigstar$}} &
            \colhead{$\delta_{nth,r}$}   &
            \colhead{$\frac{N_{e}}{10^{35}}$} &   \colhead{$T$, MK} }
\startdata
22-Apr-2001 22:37\tablenotemark{$\diamondsuit$}   & $10''\times 20''$ & $300\pm50$  & $0.7\pm0.2$  & $5\pm1$ &     $0.2\pm0.1$ &  $...$\\
03-May-2001 20:36                       & $10''\times 30''$ & $170\pm60$  & $0.25\pm01$  & $8\pm3$ &     $...$ & $5\pm2$\\
07-Aug-2001 16:25                       & $15''\times 25''$ & $600-1200$  & $3.5\pm1.5$  & $6\pm2$  &    $0.1-8$ &  $...$\\
31-Oct-2001 16:02                       & $15''\times 30''$ & $300\pm100$  & $1.2\pm0.2$  & $8\pm2$      &   $1-10$ &  $20\pm5$\\
08-Nov-2001 23:10                       & $10''\times 25''$ & $200-400$  & $1.5\pm0.5$  & $6-9$       &  $\sim10$ &  $10\pm5$\\
29-Dec-2001 16:33\tablenotemark{$\diamondsuit$}   & $20''\times 40''$ & $350\pm50$  & $1\pm0.5$  & $5-8$  &         $10-100$ &  $...$\\
30-Jan-2002 17:39\tablenotemark{$\diamondsuit$}   & $8'' \times 18''$ & $350\pm100$  & $1.5\pm0.5$  & $6\pm1$ &       $\sim80$ &  $...$\\
03-May-2002 23:32                       & $10''\times 40''$ & $550\pm100$  & $1.5\pm0.5$ (0.1)  & $5\pm1$  &   $\sim10$   & $...$ \\
21-May-2002 18:25\tablenotemark{$\heartsuit$}  & $10''\times 20''$ & $270\pm70$  & $0.3\pm0.1$ (0.03) & $7.5\pm1.5$   &  $1-3$   & $5-10$ \\
29-May-2002 17:46                       & $10''\times 20''$ & $400\pm100$  & $0.5-1.5$\tablenotemark{$\Uparrow$}(0.05)  & $7\pm1$   &  $0.1-1$  & $5-20$\tablenotemark{$\Uparrow$} \\
31-Jul-2002 23:27\tablenotemark{$\heartsuit$}  & $10''\times 30''$ & $350\pm100$  & $0.8\pm0.4$ (0.055)  & $7\pm2$  &    $0.1-3$   & $...$ \\
18-Aug-2002 20:18                       & $10''\times 30''$ & $400\pm100$  & $1.5\pm0.5$ (0.35) & $6\pm1$ &    $\sim20$  & $...$ \\
18-Aug-2002 22:28\tablenotemark{$\heartsuit$}  & $10''\times 30''$ & $200-500$  & $1-2$ (0.15) & $>7$       &   $1-10$  & $...$ \\
20-Aug-2002 22:08\tablenotemark{$\heartsuit$}  & $10''\times 12''$ & $500\pm100$  & $0.7\pm0.3$ (0.15) & $7\pm2$ &   $5\pm3$  & $...$ \\
20-Aug-2002 22:15\tablenotemark{$\heartsuit$}  & $10''\times 12''$ & $600\pm100$  & $1.2\pm0.3$ (0.015) & $7\pm2$ &   $0.1-3$   & $...$ \\
28-Oct-2002 22:51                       & $10''\times 30''$ & $500\pm100$  & $1.5\pm0.5$ (0.12) & $5\pm1$       &   $0.01-0.8$   & $...$ \\
29-Oct-2002 16:26\tablenotemark{$\heartsuit$}  & $10''\times 20''$ & $470\pm70$  & $1.2\pm0.2$ (0.4) & $5\pm1$\tablenotemark{$\sharp$}     &     $0.1-0.3$  & $...$ \\
06-Jul-2003 20:02\tablenotemark{$\heartsuit$}  & $11''\times 25''$ & $350\pm100$  & $0.6\pm0.1$ (0.11) & $5\pm1$  &     $1-10$   & $20\pm10$ \\
21-Oct-2003 20:20\tablenotemark{$\heartsuit$}  & $10''\times 30''$ & $400\pm100$  & $3\pm1$ (0.085) & $8\pm2$     &       $\sim1$   & $7\pm3$ \\
28-Oct-2003 19:05\tablenotemark{$\diamondsuit$}   & variable & $300-800$  & $0.1-2$\tablenotemark{$\Uparrow$}  & $>5$ & $0.1-1$ & $1-15$\tablenotemark{$\Uparrow$}\\
28-Oct-2003 19:46\tablenotemark{$\diamondsuit$}   & $10''\times 20''$ & $500-1000$\tablenotemark{$\cup$}  & $2\pm1$  & $5\pm1$\tablenotemark{$\cap$}\tablenotemark{$\sharp$} &     $0.3-5$  & $...$\\
\enddata
\tablenotetext{$\bigstar$}{The number density obtained from the emission measure derived from the \rhessi\ fit is shown in parenthesis when available.}
\tablenotetext{$\diamondsuit$}{No spatial information is available; typical sizes are adopted.}
\tablenotetext{$\heartsuit$}{Sizes are adopted based on \rhessi\ image.}
\tablenotetext{$\Uparrow$}{The value increases at the course of flare.}
\tablenotetext{$\cup$}{The value displays 'large-small-large' evolution pattern  at the course of flare.}
\tablenotetext{$\cap$}{The spectral index displays hardening from 12 to 5 at the rise phase and softening from 5 to 8 at the decay phase.}
\tablenotetext{$\sharp$}{The fit suggests a high-energy cutoff at an energy below 0.5~MeV.}
\end{deluxetable*}

\begin{deluxetable*}{lrrrrrr}
\tablecolumns{7}
\tablewidth{0pc}
\tablecaption{\label{T_diff_regimes}Escape time and decay constants for various trapping modes}
\tablehead{\colhead{(1)}&\colhead{(2)}&\colhead{(3)}&\colhead{(4)}&\colhead{(5)}&\colhead{(6)}&\colhead{(7)}\\
\colhead{Regime}&\colhead{$\tau(E)$; nonrel.}&\colhead{$\delta$; nonrel}&\colhead{Range of $\delta$}&\colhead{$\tau(E)$; rel.}&\colhead{$\delta$; rel}&\colhead{Range of $\delta$}}

\startdata
Weak diff.; Coulomb&      $E^{3/2}$&    $\frac{3}{2a}$&   $1<\delta<  3/2$&     $E^{1}$&    1/2&     1/2\\
Weak diff.; Turb.&        $E^{(1-\nu)/2}$&    $\frac{1-\nu}{2a}$&  $-\frac{1}{2}< \delta<0$&     $E^{(2-\nu)}$&    $\frac{2-\nu}{2}$&     $0<\delta<\frac{1}{2}$\\
Moderate diffusion&       $E^{-1/2}$&    $-\frac{1}{2a}$&     $-1/2< \delta<-1/3$&     $E^{0}$&    0&     0\\
Strong diffusion&         $E^{(\nu-3)/2}$&    $\frac{\nu-3}{2a}$&  $-1< \delta<-\frac{1}{3}$&     $E^{(\nu-2)}$&    $\frac{\nu-2}{2}$&    $-\frac{1}{2}<\delta<0$\\
\enddata
\end{deluxetable*}

\end{document}